\documentstyle[preprint,aps,graphicx,amssymb]{revtex}

\newcommand{\etal}{{\em et al.}}                
\newcommand{\beq}{\begin{equation}}
\newcommand{\eeq}{\end{equation}}
\newcommand{\bea}{\begin{eqnarray}}
\newcommand{\eea}{\end{eqnarray}}
\newcommand{\beqars}[1]{\begin{eqnarray*}{#1}}
\newcommand{\eeqars}{\end{eqnarray*}}
\newcommand{\plb}[1]{Phys.~Lett.~B {#1}}

\newcommand{\npa}[1]{Nucl.~Phys.~{A#1}}

\def\){\right) }
\def\({\left( }
\def\]{\right] }
\def\[{\left[ }

\tightenlines

\begin{document}
\draft

\title{
\begin{flushright}
{\rm MSUCL-1251}
\end{flushright}
Surface Symmetry Energy
}

\author{
Pawe\l~Danielewicz\thanks{e-mail:
danielewicz@nscl.msu.edu}
}
\address{
National Superconducting Cyclotron Laboratory and\\
Department of Physics and Astronomy, Michigan State University,
\\
East Lansing, Michigan 48824, USA\\
}
\date{\today}
\maketitle

\begin{abstract}
Binding energy of symmetric nuclear matter can be accessed
straightforwardly with the textbook mass-formula and a sample
of nuclear masses.  We show that, with a minimally modified
formula (along the lines of the droplet model),
the symmetry energy of nuclear matter can be
accessed nearly as easily.  Elementary considerations
for a macroscopic nucleus show that the surface tension needs to
depend on asymmetry.  That dependence modifies
the surface energy and
implies the emergence of asymmetry skin.  In the mass
formula, the volume and surface and (a)symmetry energies
combine as energies of
two connected capacitors, with the volume and surface
capacitances proportional to the volume and area, respectively.
The net asymmetry
partitions itself into volume and surface contributions in
proportion to the capacitances.
A combination of data on skin sizes and masses constrains
the volume symmetry parameter to 27 MeV $\lesssim \alpha
\lesssim$ 31 MeV and the volume-to-surface symmetry-parameter
ratio to 2.0 $\lesssim \alpha/\beta \lesssim$ 2.8.
In Thomas-Fermi theory, the surface asymmetry-capacitance
stems from a drop of
the symmetry energy per nucleon $S$ with density.  We
establish limits on the drop at half of normal density,
to 0.57 $\lesssim S(\rho_0/2)/S(\rho_0) \lesssim$ 0.83.  In
considering the feeding of surface by an asymmetry flux
from interior, we obtain a universal condition for
the collective asymmetry oscillations, in terms of
the asymmetry-capacitance ratio.
\\[3ex]

\end{abstract}


\narrowtext

\section{Introduction}

Changes in nuclear properties and in reaction dynamics
under the changing nucleon content of a system
attract attention in
the context
of exotic nuclear beams
\cite{bao01}.  Determination of
the density dependence of the symmetry energy, describing rise in
the nuclear energy with an
increasing asymmetry between neutrons and protons, could
permit extrapolations to
the limit of neutron matter and determinations of neutron-star
properties \cite{bao01,lat01,dan02}.
Central nuclear reactions can provide varying nuclear densities
\cite{bao01} but constraints on the symmetry energy from structure,
pertaining to normal
and subnormal densities, are also very important.  As an example,
the nuclei with a significant
excess of neutrons over protons tend to be characterized by
larger radii for the neutron than proton distributions, i.e.\
exhibit neutron skins. It has been noted, within nuclear
mean-field calculations \cite{oya98,bro00,typ01,fur02},
that the skin size could yield information on the
derivative of symmetry energy with respect to density.
For knowledge to be easily transferrable between the structure
and the reactions, which bring in their own complexity, it is important
to understand the role of symmetry energy in the structure
at a basic level.  For a utility in the reactions, importantly,
any quantitative conclusions from the structure cannot be
just tied to a specific structure model but need to pertain to
universal properties of the medium, even if coarsely.

Most elementary information on nuclear properties
stems from the semiempirical energy formula describing the
average changes in nuclear binding energy with the nucleon
content
\cite{wei35,bet36}.  The volume and surface terms in the
standard formula pertain to symmetric systems.  The volume
coefficient provides the binding energy per nucleon in
the symmetric nuclear matter; the surface coefficient, up to
a factor, provides the surface tension.  Notably, the symmetry
term in the standard energy formula has a volume character
only.
However, as we point out in this paper, basic intrinsic consistency of the
semiempirical formula requires the inclusion of a surface
symmetry term.  Quantitatively, such a term turns out further
to be essential for describing energies of light asymmetric
nuclei.  Moreover, inclusion of such a term is essential
for drawing conclusions about the symmetry energy in general,
from the nuclear structure.
The surface symmetry energy appears in the literature
in advanced energy formulas, generally in two different type of forms,
in an expanded form such as
in the liquid drop model by Myers and Swiatecki
\cite{mye66,pre75,mol95,pom02} or in more complex forms of the
liquid droplet model by the above authors \cite{mye69,mye74,mye75,mol95}.
Both formulations are currently employed in the descriptions
of nuclei \cite{mol95,pom02,sou03}.

In this work we follow a general type of considerations for
the energy of a nucleus treated as a binary macroscopic system with a surface.
Similar type of considerations can apply to
other binary systems
with an interface, such as between two phases \cite{gib48}.  We show
that, when the surface energy is affected by the particle asymmetry within the system,
then the thermodynamic
consistency requires that some of the asymmetry moves to the
surface, i.e.\ an asymmetry skin develops.
Minimization of the net nuclear energy with respect to the partition of
asymmetry produces an expression for the symmetry energy such as in the
more recent versions of the droplet model \cite{mye74,mye75}.
(In the droplet model, the skin size is a basic parameter and, as such, one of the starting points.)
We show
that the potentially confusing expression
for the energy is easily comprehended using a capacitor analogy.
The symmetry energy, quadratic in the asymmetry, can be interpreted
in terms of the energy of a capacitor for asymmetry, with the capacitance
for asymmetry proportional to the nuclear volume in the standard energy
formula \cite{wei35,bet36}.  However, in the actual nuclei, the surface
provides an additional capacitance for the asymmetry, proportional to
the surface area.  In the end, the net symmetry energy is given in terms
of the net asymmetry squared divided by the net capacitance, where one term
in the capacitance grows as the volume and the other as the area.
Notably, the liquid-drop type formula follows when the surface capacitance
is small compared to the volume capacitance.  In general, when ignoring
the surface term in the nuclear energy formula, the whole asymmetry capacitance
is attributed to the interior and, thus, the volume capacitance is overestimated
which is equivalent to underestimating the volume symmetry coefficient.

On adding a single parameter to the textbook nuclear energy formula,
representing the surface symmetry energy, we fit the measured
ground-state energies \cite{aud95} to establish limits on the symmetry energy
coefficients.  Notably, the tradition within the description of nuclear
energies in terms of advanced formulas
\cite{mye66,pre75,mol95,pom02,mye69,mye74,mye75,mol95}, often supplemented
with structure calculations,
has been to quote optimal or preferred values for utilized constants.
While this is fine as far as parameters of a model are concerned, this is not
sufficient when the fundamental matter properties are being sought.  As a drastic
example, the authors of \cite{mol95} produce parameters in two different
descriptions of nuclear energies that yield quite different surface symmetry
properties for nuclear matter, seemingly without a concern.  For the fundamental
properties, the validity of employed theory needs to be assessed and,
importantly, the features of the optimization, i.e.\ the value for the
property needs to be quoted together with an uncertainty.
However, this is normally not done in the context of advanced mass formulas.
Obviously, if the fit
minimum is very flat, or nearly degenerate, the property is not well constrained.
One plain example of how things can go wrong with a fit is when a parameter
is included that has little bearing on the quality of the fit, yet the optimization
produces a value for the parameter.  For validity of conclusions, the importance of the
specific parameter for the fit needs to be demonstrated.
For a multitude of parameters, the issues of fit stability may need to be explored.

In analyzing our fit, we demonstrate that the surface symmetry term is required,
in order to describe the energies of light asymmetric nuclei at the level similar to the
other nuclei.  However, once the surface symmetry term is included, the minimum for the fit
to masses is relatively
flat when the surface and volume symmetry coefficients are varied along a correlation
valley enforced by the need to reproduce, for heavy nuclei, the magnitude of symmetry coefficient
in the standard formula.

We subsequently turn to the analysis of the size of asymmetry skins.
The dependence of the size on net nuclear asymmetry and on
mass probes the ratio of volume to surface
symmetry coefficients.  This is because the size represents the surface asymmetry charge which,
in relation to the net asymmetry charge, tests the ratio of surface to volume
capacitances.  The dependence of skin sizes on the difference of proton-neutron
separation energies probes, on the other hand, the surface energy coefficient.
This is because, in the capacitor analogy, the difference of separation energies
represents the voltage and the dependence of surface asymmetry charge on voltage
tests the surface capacitance.
On analyzing a variety of data on skin sizes
\cite{ray79,shl79,bat89,gib92,sta94,suz95,kar02,cla02},
upon inclusion of Coulomb corrections,
we obtain constraints that, in combination with the constraints from mass analysis,
produce a relatively narrow correlation range for the volume and surface symmetry
coefficients.

In illustrating further the importance of the surface symmetry energy for nuclei,
we derive new solutions for asymmetry oscillations in nuclei, with the asymmetry flux
feeding the surface.  This extends the treatment of isoscalar modes where the internal
compression and the surface deformation are coupled \cite{boh75}.  We obtain a boundary
condition of a simple universal form, selecting wavenumbers for different multipolarities
of vibrations.
We find that the use of a realistic surface symmetry coefficient improves significantly the
description of the dependence of giant dipole resonance energy on mass, as compared to the use of
assumption that the surface accepts no asymmetry.
The mode of oscillation, in our description,
changes from one that is closer to the Goldhaber-Teller (i.e.\ surface) type
\cite{gol48} to one closer to the Steinwedel-Jensen (i.e.\ volume) type
\cite{ste50} as the nuclear mass increases.

As a final step, we analyze the nature of the surface symmetry energy at a microscopic level
within the extended
Thomas-Fermi model.  We show that an extra capacitance for the asymmetry associated
with the surface is due to the drop of
symmetry energy per nucleon with density (lower energy penalty $\equiv$ higher capacitance).
If the symmetry energy per nucleon is density independent, no extra surface capacitance emerges.
Within the Thomas-Fermi approximation, for a given density profile of symmetric matter,
the ratio of the volume to surface energy parameters
is solely related to the {shape} of
the density dependence of symmetry energy.
Equipped with this knowledge, we turn to an analysis of the mean-field structure calculations
in the literature \cite{bro00,typ01,fur02}, where numerical experimentation was used
to find a relation between the skin sizes and features of the nuclear equation of state.
The combination of our earlier results and those
in the Thomas-Fermi approximation indicates
that
the skin for a given nucleus is primarily related to the {\em shape} of the
density dependence of symmetry energy per nucleon.  Using the results from the structure
calculations \cite{fur02}, we are able to narrow the bracket on the density dependence of
symmetry energy by nearly a factor of 2 compared to what was done before \cite{fur02}.
Also, within the Thomas-Fermi
approximation, we cross-test the validity of our general results for skins.

The paper is organized as follows.  In the next section, we
discuss the symmetry energy within the binding-energy
formula and we analyze known masses in terms of the modified
formula, to constrain the symmetry energy parameters.  In Sec.\
\ref{sec:skins}, we derive expressions for
the nuclear skin size in terms of the symmetry energy parameters,
as a function of mass and charge numbers and as a function
of separation energies.
We analyze data in terms of obtained expressions to get additional
constraints on the symmetry parameters, narrowing the range allowed
by data.
The asymmetry oscillations are discussed in
Sec.\ \ref{sec:reso}.
The constraints on the low-density behavior of symmetry energy
are discussed in Sec.\
\ref{sec:micro}.  Our results are recapitulated in Sec.\
\ref{sec:conc}.

\section{Binding Energy Formula}
\label{sec:binding}

\subsection{Elementary Considerations}
We start out with a reminder of elementary issues.
The Weizs\"{a}cker nuclear energy formula\cite{wei35,bet36}
\beq
E = - a_V \, A + a_S \, A^{2/3}
+ a_a \, \frac{(N
-  Z)^2}{A}
+ a_C \, \frac{Z(Z-1)}{A^{1/3}} \, ,
\label{weiz}
\eeq
separates out different contributions to the nuclear energy.
The r.h.s.\ negative volume-term represents the energy
in the bulk limit of symmetric nuclear-matter
in the absence of Coulomb interactions.  Proportionality
to the mass number $A$ for this largest in magnitude
term, together with an approximately
constant nucleon density of
$\rho_0 = 3/(4\pi \, r_0^3) \simeq 0.160$~fm$^{-3}$ in the
nuclear interiors, underscore the short-range nature of nuclear
interactions.
The three subsequent
terms in the Weizs\"acker formula represent an increase in
the energy compared to the bulk symmetric limit.  These terms
are associated, respectively, with the size of the surface,
with an asymmetry in the number of protons $Z$ and neutrons $N$
and with the Coulomb repulsion.  Regarding the
(a)symmetry term in the Weizs\"acker formula, we observe that
this term scales like
$A$ when the nucleon number is changed and that we do not have
a surface (a)symmetry energy term, scaling like $A^{2/3}$.
The $N$-$Z$ asymmetry reduces the binding because it
enhances the role of Pauli principle and because the like-nucleon
interactions are less attractive than the neutron-proton
interactions.  The precise symmetry of that term around
$N-Z=0$ expresses the symmetry of nuclear forces under
the neutron-proton exchange, or charge symmetry.

With the radius $r_0$ of the nuclear volume per nucleon, the size of
the nuclear surface is ${\cal S} = 4 \pi \, r_0^2 \, A^{2/3}$,
and we can write the surface energy as
\beq
E_S= a_S \, A^{2/3} = \frac{a_S}{4 \pi \, r_0^2} \, {\cal S} \,
.
\eeq
The tension $\sigma$ is the derivative of energy with respect
to surface area:
\beq
\sigma = \frac{\partial E}{\partial {\cal S}}
=
\frac{a_S}{4 \pi \, r_0^2} \, .
\label{sigmadef}
\eeq
The tension represents a work per unit area needed to create
the surface.  The work is required because the nucleons at the
surface are less bound than in the interior.  However, we know
that the $N$-$Z$ asymmetry reduces the interior binding.  Thus,
with an increasing asymmetry, the work required per unit
surface,
the tension, should {\em drop}.  Coulomb effects will also affect the
surface, but will be dealt with later.

As a microscopic quantity, the tension should depend on a
microscopic quantity associated with the particle asymmetry.
The respective quantity common for different subsystems in
contact is the chemical potential.  With the nucleon chemical
potentials given by
\beq
\mu_n = \frac{\partial E}{\partial N} \hspace*{1em}\mbox{and}
 \hspace*{1em} \mu_p = \frac{\partial E}{\partial Z} \, ,
\eeq
the potential conjugate to the asymmetry $N - Z$ is half of the
the nucleon potential difference, $\mu_a = (\mu_n - \mu_p)/2$,
and
the change in energy associated with the particle numbers is:
\beq
dE = \mu_n \, dN + \mu_p \, dZ = \mu \, dA + \mu_a \, d(N-Z) \,
,
\label{eq:dE}
\eeq
with the median chemical potential $\mu = (\mu_n +
\mu_p)/2$.

Given that the tension should drop no matter, under the charge
symmetry, whether a neutron or
proton excess develops,
the tension needs to be, in the lowest order,
quadratic in the asymmetry potential $\mu_a$:
\beq
\sigma =
\sigma_0 - \gamma \, \mu_a^2 \,
\label{sigmu}
\eeq
where $\gamma >
0$.  As the tension is a derivative of the surface energy, the
dependence of $\sigma$ on $\mu_a$ implies the dependence of the
surface energy on $\mu_a$, with serious consequences.

To
demonstrate these consequences, let us construct the thermodynamic potential
generated by a Legendre transformation from the asymmetry $N-Z$
to the potential $\mu_a$:
\beq
\Phi =  \mu_a \, (N-Z) - E  \, .
\label{eq:Phi}
\eeq
For definiteness, we next consider changes in the net system when
the asymmetry changes, while
$A$ and shape remain fixed.
The derivative of $\Phi$ with respect to $\mu_a$ gives the net
asymmetry for the nucleus $N-Z$:
\beq
\frac{\partial \Phi}{\partial \mu_a}
= N - Z + \mu_a \, \frac{\partial(N-Z)}{\partial \mu_a}
- \frac{\partial E}{\partial \mu_a} = N-Z \, ,
\label{eq:dPhi}
\eeq
cf.\ (\ref{eq:dE}).  On the r.h.s.\ of (\ref{eq:Phi}), we can separate
the energy $E$ into the surface $E_S$ and volume $E_V$ contributions,
$E=E_S+E_V$.  When the asymmetry changes, the change in volume energy
is $dE_V = \mu_a \, d(N_V-Z_V)$, where $d(N_V -
Z_V)$ is the change in particle asymmetry within the nuclear
volume, expected equal to $d(N-Z)$.  If we now substitute
(\ref{eq:Phi}) into the l.h.s.\ of (\ref{eq:dPhi}) and compare to the
r.h.s.\ there, we find
\beq
\mu_a \, \left(\frac{\partial (N - Z)}{\partial\, \mu_a}
- \frac{\partial (N_V - Z_V)}{\partial\, \mu_a} \right) -
\frac{\partial E_S}{\partial \, \mu_a} = 0 \, .
\eeq
Here we can see, though, that we will run into a contradiction if we identify
changes in the net nuclear asymmetry $N-Z$ with the changes in the
interior asymmetry $N_V-Z_V$, when the surface energy $E_S$ depends on chemical
potential $\mu_a$.  The net and interior asymmetries must be different, and the
difference $N_S-Z_S = N-Z - (N_V-Z_V)$ must, as a consequence, reside within
the nuclear surface.  To tackle the conceptual issue of how
a surface can contain a particle number, we must consider how
the surface quantities get to be defined.

\begin{figure}

\centerline{\includegraphics[angle=0,
width=.70\linewidth]{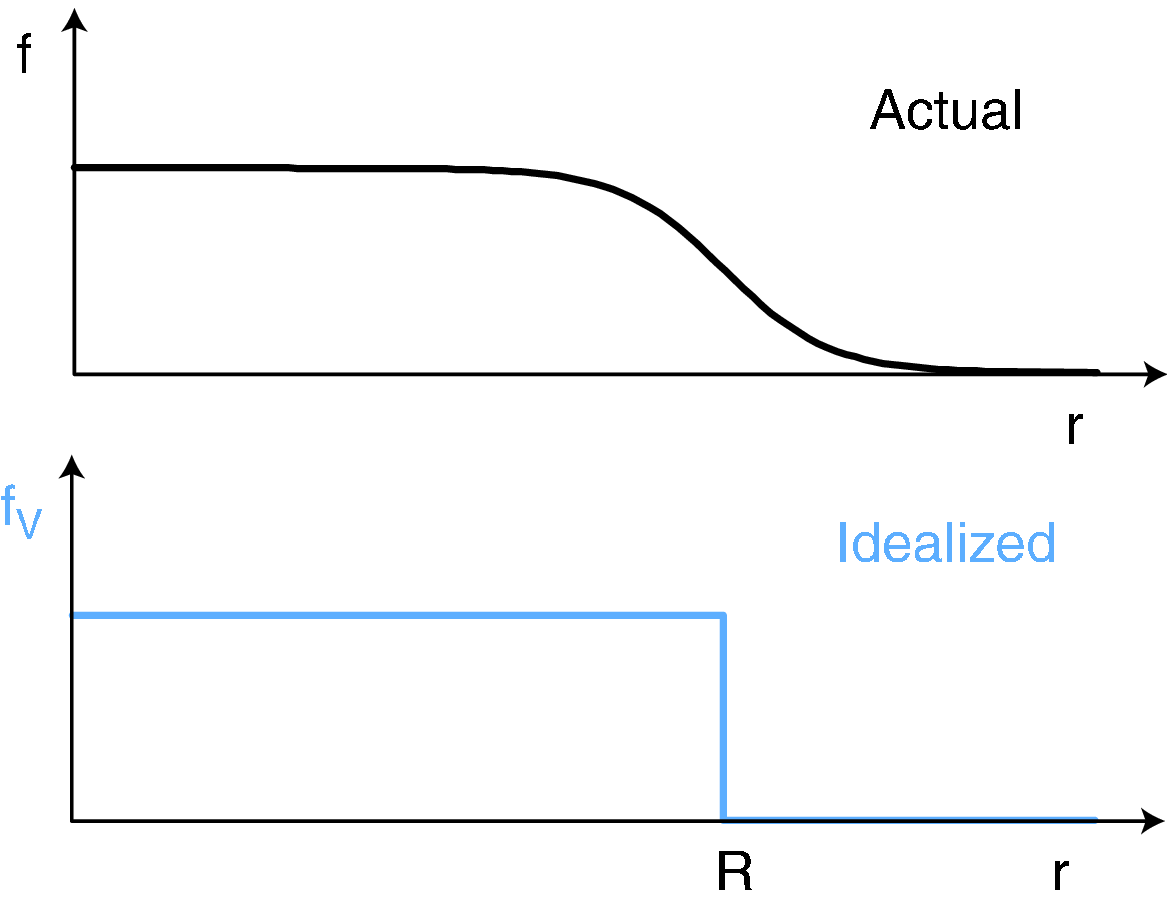}}
\vspace*{.2in}
\caption{
Illustration for the construction of surface quantities
following Gibbs\protect\cite{gib48}.  The top shows the
variation of the density $f$
for some quantity $F$ within the system.  The bottom shows an
idealized reference system, where the density changes in a
discontinuous manner, right at the surface, between the density
values
$f_V$ characteristic for the system away from the interface.
The surface quantities $F_S$ are defined as the
difference
between $F$, for a macroscopic volume containing the interface,
and $F_{id}$, for the same volume in the reference system, $F_S
= F - F_{id}$. }
\label{fig:gibbs}
\end{figure}

For an interface between phases in equilibrium, Gibbs\cite{gib48}
suggested to take a small but macroscopic volume surrounding an
element of the interface and, besides the actual system, consider
an idealized reference system, cf.\ Fig.\ \ref{fig:gibbs}.  In the
idealized system, the densities of thermodynamic quantities, such
as the particle density, change abruptly at the interface from a
value characterizing one of the phases away from the interface to
the value for the other phase.  The surface share $F_S$ for the
surface element ${\mathcal S}$ is then the difference between the
value of the quantity for an actual system $F$ and the value for
an idealized system $F_{id}$: $F_S = F-F_{id}$.  In terms of
densities, for the case of a nucleus in equilibrium with the
vacuum, this further yields for a quantity~$F$: $F_S = {\mathsf f}
\, {\cal S} = F - V \, f_{V}$, where $f_V$ is the density of $F$
characteristic for the interior and ${\mathsf f}$ represents the
surface density.  The problem is, however, that the result for the
surface depends on a precise position assumed for the interface,
and a condition may be imposed further to specify the position.
For a nucleus it is natural to assume that the surface contains no
mass number and, for a spherical nucleus, this is equivalent to
taking the equivalent sharp-sphere radius for the surface of
$R=r_0 \, A^{1/3}$.

\begin{figure}

\centerline{\includegraphics[angle=0,
width=.70\linewidth]{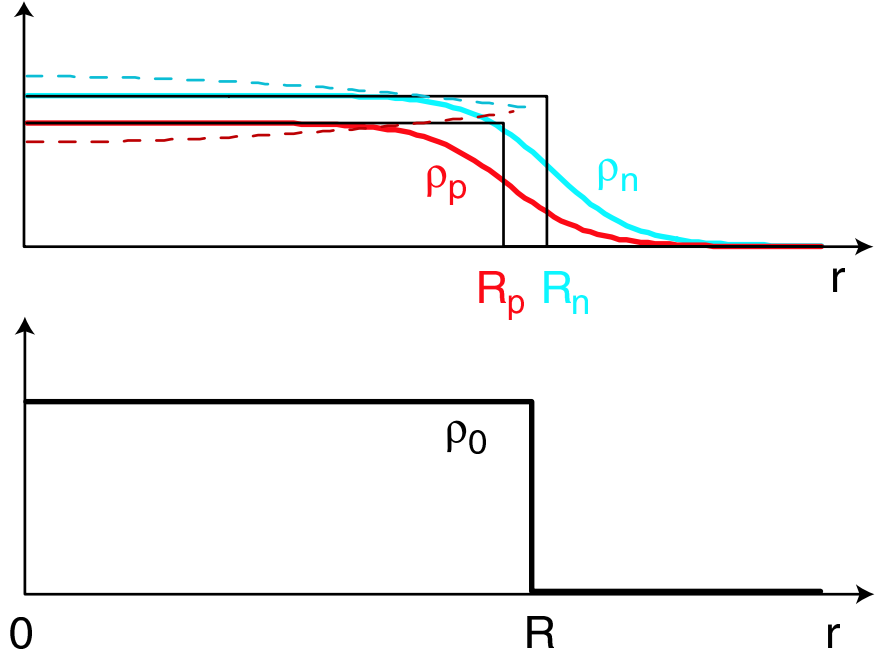}}
\vspace*{.2in}
\caption{
In a two-component system, the surface-attributable
particle-asymmetry
can arise even when the surface is
constructed to have no net particle number.  For a spherical
ground-state nucleus,
this corresponds to the situation when the sharp-sphere
equivalent proton $R_p$ and neutron $R_n$ radii differ from
each other and bracket the matter sharp-sphere radius $R$.
The dashed lines in the top part of the figure illustrate
the polarization of a nuclear interior, in the form of a
sharp-edged sphere of radius $R$, by the Coulomb interactions.
}
\label{fig:gibbs2}
\end{figure}

Only one condition related to the particle numbers, however,
suffices to localize the surface.  In a two-component system,
the attempts to localize the surface using one of the
components or
the other would lead to two different positions.  If the
surface is localized using the net particle density, a finite
surface asymmetry $N_S-Z_S$ results, even though the surface
has a vanishing net nucleon number
$A_S =
N_S + Z_S=0$.
For a spherical nucleus, this corresponds to
having two different equivalent sharp-sphere radii for neutrons
$R_n$ and protons $R_p$, bracketing the sharp-sphere radius for
the matter $R$, cf.\ Fig.\ \ref{fig:gibbs2}.

The Gibbs prescription can be used for clarifications, but also
as a direct tool, as e.g.\ later illustrated by a consideration in
Subsection \ref{ssec:symTF} or when pursuing curvature effects.
We next relate the surface asymmetry excess to the asymmetry
potential.
Either by following the Gibbs prescription, with a small volume
surrounding some surface element, or by isolating contributions
pertaining to the surface
element in the energy
considerations before, we find
\beq
d \, E_S = \sigma \, d{\cal
S} + \mu_a \, d(N_S - Z_S) \, ,
\label{eq:des}
\eeq
under independent variations of asymmetry and of the surface
element size.
From changes in $E_S$, when
taking an increasing fraction of the net surface, we find that
\beq
E_S = \sigma \, {\cal S} + \mu_a \, (N_S - Z_S) \, . \label{ES}
\eeq
On taking a differential of the above, with the result
\beq
dE_S = {\mathcal S} \, d\sigma + \sigma \, d{\mathcal S}
+(N_S - Z_S) \, d\mu_a + \mu_a \, d (N_S-Z_S) \, ,
\eeq
and on subtracting from this result the
relation (\ref{eq:des}), side by side, we obtain
a Gibbs-Duhem type relation.  That last relation
yields
\beq
\frac{N_S - Z_S}{\mathcal S} = - \frac{d\,
\sigma}{d\, \mu_a} = 2 \, \gamma \, \mu_a \, ,
\label{nzmu}
\eeq
where, at the end, we employed Eq.\ (\ref{sigmu}).

With (\ref{nzmu}) and (\ref{sigmu}), we get from (\ref{ES}):
\beq
E_S = \sigma_0 \, {\cal S} + \gamma \, \mu_a^2 \, {\cal S} =
E_S^0 + \frac{1}{4 \gamma} \, \frac{(N_S - Z_S)^2}{\cal S}
\, ,
\label{ESnz1}
\eeq
where $E_S^0$ is the surface energy in absence of the surface
asymmetry.
At this point, we may cross-check the consistency of our result
(\ref{ESnz1}) with its derivation.  Thus, from (\ref{ESnz1}), we
find
\bea
\nonumber
dE_S & = & dE_S^0 + \frac{1}{4\gamma} \, \frac{2(N_S-Z_S)}{\mathcal S} \,
d(N_S-Z_S) - \frac{1}{4\gamma} \, \frac{(N_S - Z_S)^2}{{\mathcal S}^2}
\, d{\mathcal S} \\
& = & (\sigma_0 - \gamma \, \mu_a^2) \, d{\mathcal S} + \mu_a \,
d(N_S -Z_S) = \sigma \, d{\mathcal S} + \mu_a \,
d(N_S -Z_S) \, ,
\eea
which agrees with (\ref{eq:des}) and where we utilized (\ref{nzmu})
and $dE_S^0 = \sigma_0 \, d{\mathcal S}$.  In connection with (\ref{ESnz1}),
we may note that although the surface tension decreases with the asymmetry,
Eq.\ (\ref{sigmu}), the net surface energy increases with the
asymmetry.

On account of charge symmetry for the interior nuclear energy per unit volume,
to lowest order in the asymmetry, the volume energy will
have the form
\beq
E_V=E_V^0 + V \, \,  \alpha \, \rho_0 \left( \frac{N_V - Z_V}{A} \right)^2  =
E_V^0 + \alpha \, \frac{(N_V - Z_V)^2}{A} \, ,
\label{EVnz}
\eeq
where we factorized the coefficient in the energy density so as to get
the net volume energy in conformance with (\ref{weiz}) and where
$E_V^0$ ($\equiv a_V \, A$) is the energy for a symmetric system.
For an analogy with (\ref{EVnz}), we introduce
a constant with the dimension of energy
in $E_S$, $\beta= 1/(16\, \pi \, r_0^2 \, \gamma)$, to write the
r.h.s.\ of (\ref{ESnz1})
as
\beq
E_S = E_S^0 + \beta \, \frac{(N_S - Z_S)^2}{A^{2/3}} \, .
\label{ESnz}
\eeq
In the volume energy we use a constant $\alpha$ in place of $a_a$ of
(\ref{weiz}) to make
a distinction between the symmetry constant in the bulk and the
constant in use in the binding-energy formula.  As will be next
seen, $a_a$ in use generally results from a combination of
$\alpha$ and $\beta$.

Minimization of the net system energy $E$ under the constraint of net
asymmetry,
\beq
E = E_V + E_S
\hspace*{2em}\mbox{with}\hspace*{2em} N - Z = N_V - Z_V + N_S -
Z_S \,  ,
\label{ENZ}
\eeq
yields the condition of equal chemical
asymmetry potentials for the interior and surface,
\bea
0 &=& \frac{\partial E}{\partial (N_V-Z_V)}
= \frac{\partial E_V}{\partial (N_V-Z_V)}
- \frac{\partial E_S}{\partial (N_S-Z_S)} \nonumber \\[.5ex]
& = & 2\alpha \, \frac{N_V-Z_V}{A} - 2\beta \, \frac{N_S-Z_S}{A^{2/3}} \, .
\eea
With this, we find for the asymmetry ratios
\beq
\frac{N_S - Z_S}{N_V - Z_V} =
\frac{\alpha}{\beta \, A^{1/3}} \, ,
\hspace*{2em}
\mbox{
and
}
\hspace*{2em}
\frac{N_S - Z_S}{N - Z} = \frac{\frac{N_S - Z_S}{N_V - Z_V}}{1+\frac{N_S - Z_S}{N_V - Z_V}}
= \frac{1}{1+ \frac{\beta }{\alpha}\, A^{1/3}} \, .
\eeq
The net energy then becomes equal to
\bea
\nonumber
E & = & E_V + E_S = E_0 + \frac{\alpha}{A} \, \frac{(N-Z)^2}{(1+ \frac{\alpha}{\beta }\, A^{-1/3})^2}
+ \frac{\beta}{A^{2/3}} \, \frac{(N-Z)^2}{(1+ \frac{\beta }{\alpha}\, A^{1/3})^2} \\
& = & E_0 + \frac{\alpha}{1 +
\frac{\alpha}{\beta} \, A^{-1/3}} \, \frac{(N-Z)^2}{A} \, .
\label{EVSNZ1}
\eea

\subsection{Capacitors for Asymmetry}

If a problem can be mapped onto one that is well understood, existing intuitions can be
exploited to make predictions and lay out results with little need for any derivations,
and this pertains to the last results.  The quadratic
energies (\ref{EVnz}) and (\ref{ESnz}), with the constraint (\ref{ENZ}), invite e.g.\ an analogy
to two connected springs, of spring constants $\frac{\alpha}{2A}$
and $\frac{\beta}{2A^{2/3}}$, respectively,
stretched between separated supports.  The context, however, suggests an even more appropriate analogy
to two connected capacitors carrying charges $q_X=N_X-Z_X$ ($X=V,S$), of capacitances $C_V=\frac{A}{2\alpha}$ and
$C_S=\frac{A^{2/3}}{2\beta}$, respectively, to yield capacitor energies $E_X= E_X^0+\frac{q_X^2}{2C_X}$.
As the asymmetry moves
in-between the capacitors, the equilibrium situation is that of a charge in proportion
to capacitance, yielding
\beq
\frac{N_S - Z_S}{N_V - Z_V} = \frac{C_S}{C_V}= \frac{\alpha}{\beta \, A^{1/3}}
\hspace*{2em}
\mbox{
and
}
\hspace*{2em}
\frac{N_S - Z_S}{N - Z} = \frac{C_S}{C_V+C_S}= \frac{1}{1+ \frac{\beta }{\alpha}\, A^{1/3}} \, .
\label{NZfrac}
\eeq
The net capacitor energy, finally, is given in terms of the net charge and net capacitance
\beq
E=E^0+\frac{q^2}{2(C_V+C_S)}= E^0+ \frac{(N-Z)^2}{\frac{A}{\alpha}+\frac{A^{2/3}}{\beta}}
= E_0 + \frac{\alpha}{1 +
\frac{\alpha}{\beta} \, A^{-1/3}} \, \frac{(N-Z)^2}{A} \, .
\label{EVSNZ}
\eeq
The connected capacitor and the nuclear symmetry energy problems are,
at this level, completely equivalent.

The symmetry energy of (\ref{EVSNZ}) should, generally, replace
that in the energy formula (\ref{weiz}), with $\beta \rightarrow
\infty$ ($\gamma \rightarrow 0$, in (\ref{sigmu})) obviously
corresponding to the standard formula. In (\ref{NZfrac}), we
observe that the surface fraction of the asymmetry excess
decreases as nuclear mass increases, because the surface
capacitance grows only as $A^{2/3}$.  It may be tempting to expand
the symmetry energy (\ref{EVSNZ}) in the surface-to-volume
capacitance fraction.  However,
whether this can be done depends on the volume-to-surface
coefficient ratio $\alpha/\beta$. For large enough ratios and
small $A$, we, in fact, principally might be able to ignore 1 in
the denominator and get the symmetry energy such as characteristic
for the surface storage of asymmetry. (This would express the fact
that, just as the volume capacitance grows faster than surface
capacitance, it also drops faster.) Later in this and in the
following section we shall see what $\alpha/\beta$ values are
favored by data.

\subsection{Coulomb Effects}

So far, we have not considered the effect of Coulomb interactions
on the surface asymmetry and, generally, the interplay
of Coulomb and symmetry energies
in the net energy. The Coulomb energy may be
represented as
\bea
E_C & = & E_V^C + E_{VS}^C + E_S^C =
\frac{e^2}{4 \pi  \epsilon_0} \, \frac{1}{R} \left( \frac{3}{5}
\, Z_V^2 + Z_V \, Z_S + \frac{1}{2}\, Z_S^2\right) \nonumber
\\
& = & \frac{a_C}{A^{1/3}}
\left(Z_V^2 + \frac{5}{3} \, Z_V \, Z_S + \frac{5}{6}\, Z_S^2
\right) \,
\label{ECVS}
\eea
where the three terms correspond to the volume, volume-surface
and surface interactions, respectively, and
$Z_V=Z-Z_S$.

The combination of the symmetry energies
(\ref{EVnz}) and (\ref{ESnz})
and the Coulomb energy (\ref{ECVS}) is quadratic in the surface
excess and is easily found to minimize at:
\beq
N_S - Z_S
=
\frac{N-Z - \frac{a_C }{12 \,
\alpha}\, Z \, A^{2/3}}{1+ \frac{\beta }{\alpha}\, A^{1/3} +
\frac{a_C }{48 \, \alpha}\, A^{2/3}}
\approx
\frac{N-Z - \frac{a_C}{12 \,
\alpha} \, Z \, A^{2/3}}{1+ \frac{\beta}{\alpha} \, A^{1/3}} \, .
\label{nzs}
\eeq
The last approximation expresses the fact the Coulomb
interactions represent a very soft spring in the energy
combination,
compared to the volume asymmetry spring.  Compared to
(\ref{NZfrac}), we see that, as might be expected, the Coulomb
interactions try to tilt the surface asymmetry towards a proton
excess.
Quantitatively, however, the Coulomb correction to the
asymmetry excess is quite negligible for light and medium
nuclei and acquires only a limited significance for heavy
nuclei.

As to the energies, even when the Coulomb correction to the
surface
asymmetry has a significance, the corresponding changes in the
asymmetry energy are quite ignorable.  This because around
the minimum, specified by (\ref{NZfrac}), the changes in the
asymmetry energy
are quadratic in the asymmetry correction.  On the other hand,
the correction to the Coulomb energy, from the development
of surface asymmetry,
\beq
\Delta E_C \simeq \left. \frac{\partial E_C}{\partial Z_S}
\right|_0 \, Z_S  \simeq \frac{a_C \, Z^2}{A^{1/3}} \,
\frac{N-Z}{6 \, Z} \, \frac{1}{1+ \frac{\beta}{\alpha} \,
A^{1/3}} \, ,
\label{DEC}
\eeq
may have a moderate significance within the net energy and, in heavy
nuclei, can reach a comparable magnitude to the correction to
the Coulomb energy from surface diffuseness.

\subsection{Relation to the Droplet Model}

The droplet model \cite{mye69} utilizes the relative displacement
of neutron and proton surfaces as one of the different parameters describing
a nucleus.  A global energy functional is constructed for a nucleus,
incorporating an effect of the surface displacement on the net energy.
This functional is minimized with respect to the different parameters in order to
arrive at the ground-state energy.

While the droplet model contains a variety of potentially useful details, these can
easily shroud the issues of symmetry energy.  In fact, our own understanding
of that aspect of the model stems largely from a mapping of the
results there onto those in the present paper.  In particular,
if the final result for energy in the droplet model in \cite{mye74} (and in
later papers on the model)
is expanded in the net particle asymmetry, then a symmetry energy emerges such as
in Eq.\ (\ref{EVSNZ1}),
provided our constants are identified with those in the droplet model as:
\beq
\alpha = J \hspace*{2em}\mbox{and}
\hspace*{2em} \beta = \frac{4}{9} Q =
\frac{\frac{4}{9} H}{1 - \frac{2}{3} \, P/J} \, .
\label{dropletID}
\eeq
The individual constants $H$, $P$ and $G$, found equal to $G=3J\, P/2Q$,
describe the assumed dependence \cite{mye69} of the surface energy
on bulk asymmetry and on normalized size of the asymmetry skin.
The different definitions of the surface energy within the model are
discussed in Refs.\ \cite{mye85,mye01}.

If the net energy in the droplet model \cite{mye74,mol95} is expanded
simultaneously in terms of
the net asymmetry and in the Coulomb coefficient $a_C$, then
a correction to the energy is found in the form such as
in (\ref{DEC}).

\subsection{Comparison to Binding-Energy Data}

We next examine what can be learnt on the nuclear symmetry energy from
measured binding energies \cite{aud95}, $B=-E$, employing the results from above,
by following an energy formula with only one more adjustable
parameter than in the textbooks \cite{boh75},
to represent the surface symmetry energy.
Specifically, we fit the measured energies with
\beq
E = - a_V \, A + a_S \, A^{2/3}
+
 \frac{\alpha}{1 +
\frac{\alpha}{\beta} \,A^{-1/3}} \, \frac{(N-Z)^2}{A}
+ a_C \, \frac{Z \, (Z-1)}{A^{1/3} \, (1+\Delta)} - \delta \, .
\label{weiz2}
\eeq
Compared to (\ref{weiz}), we changed the symmetry term to
(\ref{EVSNZ}) and we added the standard pairing term $\delta =
\pm
a_p \, A^{-1/2}, 0$.  We either carry out fits with $\Delta=0$
in the Coulomb term, i.e.\ the energy formula modified, from
the standard, only by the inclusion of surface symmetry
energy, or with
\beq
\Delta = \frac{5 \, \pi^2}{6} \, \frac{d^2}{r_0^2 \, A^{2/3}}
-
\frac{1}{1+ \frac{\beta
}{\alpha} \, A^{1/3}}
\,
\frac{N-Z}{6 \, Z}
\, ,
\label{delta}
\eeq
including the effects of surface diffuseness and of the
asymmetry skin (\ref{DEC}) on the Coulomb energy.  In the
first r.h.s.\ term above, $d \approx 0.55$~fm is the diffuseness
parameter in the
Fermi function from the parametrization of nuclear charge
distributions \cite{jag74},
\beq
\rho(r)=\frac{\rho_0}{1+\exp{[({r-c})/{d}]}} \, ;
\label{rhor}
\eeq
that term
stems from an expansion of the Coulomb energy for (\ref{rhor})
in terms of $d$.

Qualitatively, results from the two fits indicated above are similar
but they differ in quantitative details.  Obviously, inclusion
of the volume and surface competition, in the symmetry energy,
changes the dependence of the net energy on asymmetry as a
function of mass.  However, the Coulomb energy is also a source of
the dependence on asymmetry and deciding on the details in one energy
contribution depends on the details in the other.  We believe that
more reliable results are obtained with the finite $\Delta$, Eq.\
(\ref{delta}), in (\ref{weiz2}) as we only incorporate information
available from electron scattering or following from intrinsic
consistency of the considerations.  The number of independent
parameters in the fit with (\ref{delta}) is the same as in the
$\Delta=0$ fit, precluding, in practice, a fit runaway in
the parameter space.

In the fits, we optimize the sum of absolute
deviations of the theoretical energies from experimental,
rather than the sum of the deviations squared, as the intention is
to provide an average description of the nuclear features.
Results, however, weakly depend on that choice.
In the optimization, we include the energies of all nuclei
with $A \ge 2$, but exclude the energies deduced from trends
\cite{aud95}.
We either let
all parameters in (\ref{weiz2}) vary in an unconstrained manner or we constrain
$\alpha/\beta$ or both of the symmetry parameters.

\begin{figure}

\centerline{\includegraphics[angle=0,
width=.70\linewidth]{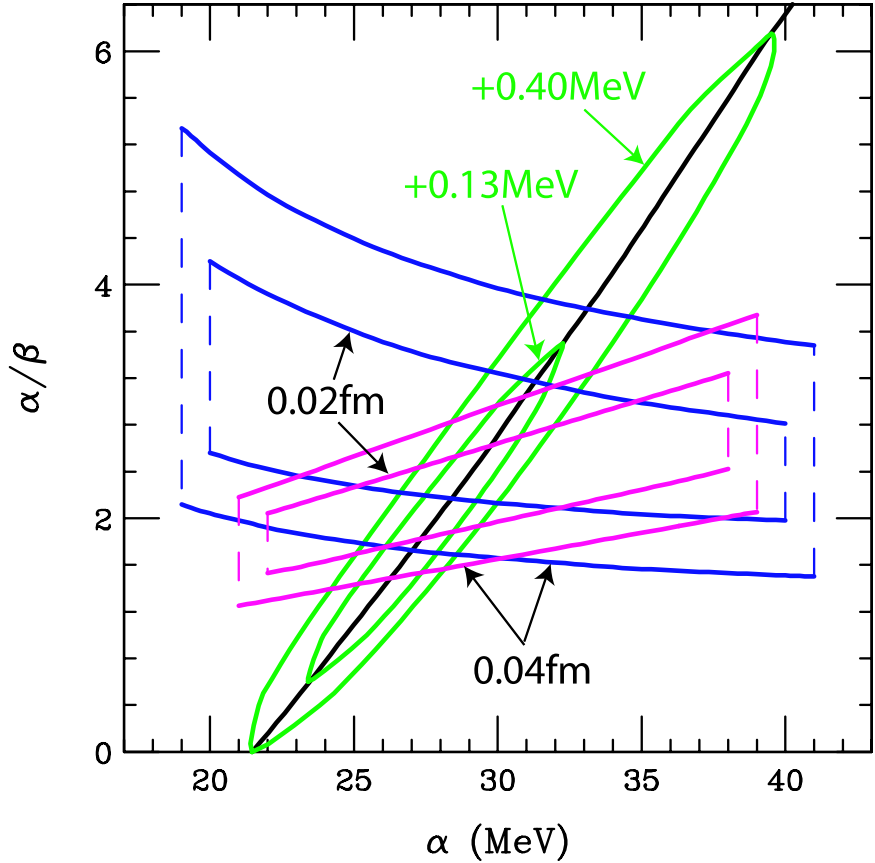}}
\vspace*{.2in}
\caption{
Results from
optimizations of the volume parameter $\alpha$
and the volume-to-surface parameter ratio $\alpha/\beta$. The
thick solid diagonal line shows the optimal values of $\alpha$
at fixed $\alpha/\beta$ when fitting the nuclear masses.
The oval contour lines show the loci of parameter
combinations
that yield average absolute deviations from the measured nuclear
masses larger by 0.13 and 0.40 MeV, respectively, than a
best-fit
parameter combination.  The rectangular contour
lines indicate the loci of
parameter combinations at constant $\chi^2$ when fitting the
asymmetry skins deduced from data, either following Eq.\ (\ref{rmsnp}) or (\ref{rmsnpmu}).
Only the solid more horizontal
portions of the skin-thickness contours are meaningful -- the vertical
dashed portions
serve only to guide the
eye.
The loci are for the
$\chi^2$-values assuming a combination of the standard
statistical
deviation and of the average systematic error of a skin-size
formula of
0.02 and 0.04~fm, respectively, with the solid positively sloped contours
representing Eq.\ (\ref{rmsnpmu}) and the negatively sloped representing Eq.\ (\ref{rmsnp}).
}
\label{fig:rabc}
\end{figure}

The average per nucleus deviation of the theoretical energy from experimental
exhibits a narrow valley in the space of $\alpha/\beta$ vs
$\alpha$, as illustrated in Fig.\ \ref{fig:rabc} for the fit
with (\ref{delta}).  The valley is described by the equation
$\alpha \approx
(21.5 + 3.1 \, \alpha/\beta)$~MeV.
An analogous valley is found for the fit
with
$\Delta=0$, but the valley is then shifted by about 10\% to higher
values in $\alpha$.  The strong correlation may be understood
in terms of the need to properly describe energies for the
heaviest
nuclei with largest asymmetries requiring $\alpha/(1 +
A^{-1/3} \,\alpha/\beta ) \approx a_a$, for $A \sim 200$.

Along the valley, the average deviation minimizes at
$\alpha/\beta \sim 1.7$, with the value of $\overline{|E-E^{exp}|} = 1.97$~MeV.
However, the minimum along the valley
is rather flat.  To understand the
nature of the minimum, we carry out the following procedure.
We carry out an optimization of all parameters in the energy
formula at a fixed $\alpha/\beta$.  Then, keeping all
parameters fixed including $\alpha/\beta$ but excluding
$\alpha$, we invert the energy formula to deduce $\alpha$ from
the mass of an individual nucleus.
With this, we test locally
the global fit.  If the formula provides a proper description,
then, on the average, we should get the same $\alpha$ from an
individual mass as from the global fit.
Notably,
for the inversion to be significant,
the asymmetry should be large and we demand that $|N-Z|/A > 0.2
$.  Sample results are displayed in Fig.\
\ref{fig:alphad}, as a function of $A$.

First to consider are the results obtained at $\alpha/\beta=0$,
i.e.\ ignoring the surface symmetry energy, as in the textbook
mass formulas.
Represented by the diamonds in Fig.\ \ref{fig:alphad},
the results for $\alpha$ (i.e.\ $a_a$ in this
case)
group coarsely
into a line.  At large $A$, the line from symbols oscillates,
exhibiting the shell
structure, around the value of $\alpha$ from the global fit,
represented by a straight line.  However, at $A \lesssim 50$
the values from a local inversion systematically drop below $a_a$ from
the global fit.  Such a drop
for $a_a$ at low $A$ is expected
from a competition between the volume and surface symmetry
energies (\ref{weiz2}).  If we raise the parameter ratio, such
as to the value of $\alpha/\beta = 2$, we can get the locally
restored values of $\alpha$ to oscillate around the global
fit value down to very low~$A$.  Regarding the restored values of
$\alpha$, we may note in Fig.\ \ref{fig:alphad} that they
locally exhibit less scatter
for the fit with (\ref{delta}) than with $\Delta = 0$, suggesting
that the first fit grasps the physics better.  Finally, if we
assume a too high value of the parameter ratio, such as
$\alpha/\beta=6$, we get the restored $\alpha$ to rise up at
low $A$ in an opposite tendency to that for $\alpha/\beta=0$.

The gain in the description of low-$A$ nuclei when using a finite $\alpha/\beta$
can be further
appreciated by looking directly at magnitudes of the average fit
deviations in the low-$A$ region.
Thus, if we employ $\alpha/\beta=0$ in (\ref{weiz2}) in the
global fit with (\ref{delta}), then for nuclei in the region of $5 \le A \le 50$ and
$|N-Z|/A>0.2$, we get $\overline{|E-E^{exp}|} = 7.4$~MeV, far in
excess of the average deviation of 2.37~MeV for all nuclei.
On the other hand, for $\alpha/\beta \sim 1.7$, we get an average
deviation of $\overline{|E-E^{exp}|} = 2.3$~MeV in the region of
$5 \le A \le 50$ and $|N-Z|/A>0.2$, close to the average deviation of 1.97~MeV
for all nuclei.

While, at the lower $A$, the use of an intermediate value of
$\alpha/\beta$ improves
the performance of the energy formula, at the
highest
$A$ the separate sensitivity to the ratio and $\alpha$ is quite reduced.
With most of the measured nuclei of high anisotropies at the
high $A$,
the reduction in the deviation
between the
formula and the experimental energies, when averaged over {all}
$A$, is just $\sim 0.40$~MeV per
nucleus for
the absolute minimum
in Fig.~\ref{fig:rabc},
compared to the optimal case of $\alpha/\beta=0$.
Considering in particular Fig.\
\ref{fig:alphad}, we need to decide how accurate our
phenomenological theory should be in describing the nuclei.
While the superiority of the $\alpha/\beta=2$ case over the
other cases in Fig.\ \ref{fig:alphad} is quite apparent, cases
that might
be less than a third of the way in-between $\alpha/\beta=2$ and
the two other would not be clearly superior, given the scatter
of the points and the lack of shell-effect description in
the theory.  With this, we consider the locus of points in the
$(\alpha,\alpha/\beta)$ plane corresponding the deviation
reduced by a factor of 1/3 compared to the optimal
$\alpha/\beta=0$ point
(i.e.\ 0.13 MeV in Fig.~\ref{fig:rabc}), as the approximate
boundary of the parameter
range favored by the energy formula.  Have we had a theory that
could e.g.\ predict the shell effects, we might have been able
to impose a more stringent limit on the parameters.
For reference, we further
show in Fig.\ \ref{fig:rabc} the boundary for the 0.40~MeV
deviation, passing
through the optimal $\alpha/\beta=0$ point and in the vicinity
of the optimal
$\alpha/\beta=6$ point.

On one hand, the number of parameters here, in accessing the symmetry energy
from the binding-energy data, represents a bare minimum, with the
results necessarily being somewhat coarse.  On the other hand, there can be
advantages to the fits with a low parameter-number, besides
the obvious immediate accessibility of the analysis.  While
coarse, the low parameter-number fits tend to be stable and faithful, without
the optimization being thrown off by obscure features of data
or by auxiliary theoretical inputs.  As an example of dangers
encountered when increasing the parameter number, when making the
surface diffuseness $d$ in (\ref{delta}) a variable parameter \cite{pom02},
we find that the average deviation from measured energies minimizes for sensible
parameter values.  However, that minimum is found to be very flat when the diffuseness
and symmetry parameters are varied simultaneously, extending far out
into the region of large surface-to-volume parameter ratios and large
diffuseness.
Such behavior of a minimum marginalizes the significance
of optimal values for a fit, given that, even when experimental errors are
small, the theoretical uncertainties need to be incorporated \cite{tar87}.
The situation can be remedied by bringing
in external knowledge \cite{tik63}, such as from electron scattering in the
case in question.  Such issues may be obvious for experimentalists,
but are not routinely appreciated by theorists.

\section{Asymmetry Skins}
\label{sec:skins}

\subsection{Interest in Skins}
Interest in
asymmetry skins has surged
in the
context of
the general interest in asymmetry effects, stemming from
the availability
of radioactive beams.
Expectations were raised that an analysis of
the asymmetry skins could permit
to extrapolate the nuclear matter properties
to the neutron matter \cite{oya98,bro00,typ01,fur02}.
Different features of effective interactions have been tested with
regard to their bearing on the thickness of the skins.
In several works\cite{hor02}, a
direct connection between the neutron skin in the
spherical $^{208}$Pb nucleus and the neutron star structure was
made.
Primarily, in recent times, the skins have been theoretically
investigated within the microscopic structure calculations
\cite{miz00,vre00,ner02}.

Unfortunately,
experimental extraction of the neutron distributions has been
difficult.  Different methods have been employed, such as
analyses of nucleon scattering
\cite{ray79,shl79,kar02,cla02} (including the analysis of
polarized proton scattering and of neutron
and proton scattering) or analyses combining information from
strongly interacting ($p$, $\pi^\pm$, $\alpha$, $^{12}$C) and
electromagnetic probes ($e$ scattering, isotopic transition
shifts) \cite{bat89,gib92,sta94,suz95}.  Recently, it has been
proposed \cite{hor01} to utilize parity violations in electron
scattering to determine neutron rms radii, contributing to the
surge of interest in the asymmetry skins.

\subsection{Skin Size}
To assess the thickness of the asymmetry skin for a given surface
excess (\ref{nzs}), let us consider the sharp sphere radii for
neutrons $R_n$ and protons $R_p$.  These radii correspond to
volumes with interior densities such as for the bulk matter, i.e.\
\beq N = \frac{N_V}{A} \, \rho_0 \, \frac{4}{3}\pi  R_n^3 = N_V
\left(\frac{R_n}{R}\right)^3 \simeq N_V  \left(1 + \frac{3(R_n
-R)}{R}\right) \, . \label{N=} \eeq In the last step, we use the
fact that the intrinsic consistency of our considerations requires
that we can tell the surface from the interior and, thus, must have
$|R_n - R_p| \ll R$ or, equivalently, $|N_S-Z_S| \ll A$. We next
get from (\ref{N=}) \beq \frac{R_n - R}{R} = \frac{N_S}{3  N} \, .
\eeq With an analogous result for $R_p$, we find for the
difference of the sharp-sphere radii:
\beq
\frac{R_n - R_p}{R} =
\frac{A \, (N_S - Z_S)}{6  N  Z} = \frac{A}{6  N  Z} \, \frac{N-Z
- \frac{a_C}{12 \, \alpha} \, Z \, A^{2/3}}{1+
\frac{\beta}{\alpha} \, A^{1/3}} \, ,
\label{rnp}
\eeq
where we inserted the result from (\ref{nzs}).
We see that
the difference between the neutron and proton radii is primarily
linear in the asymmetry and measures the symmetry coefficient
ratio $\alpha/\beta$. For a surface with no capacitance for the
asymmetry, characterized by $\alpha/\beta=0$, the asymmetry skin
should vanish.  If one were trying to compare the result (\ref{rnp}) with
the results in \cite{mye69}, one should note that there, in derivations,
the approximation of $N\approx Z \approx A/2$ is utilized.

Regarding (\ref{rnp}), the results of measurements
and of microscopic
calculations for the skins are usually, however, not expressed in terms
of the neutron-proton difference of sharp-sphere radii but rather
the difference of rms radii.
The difference between the rms compared to the sharp-sphere radii,
aside from the standard reduction factor
(dropping out from the ratio of radii) and from a correction
for diffuseness, brings in
a Coulomb correction additional to that already in (\ref{rnp}).
That additional correction stems from a polarization of the nuclear
interior by
the Coulomb forces.  The polarization is schematically
indicated in Fig.\ \ref{fig:gibbs2}.
In order to assess the polarization effect, within the
net nuclear energy we represent the
interior symmetry and
Coulomb energies in integral forms.
We expand these energies,
respectively,
to the second and first order in the deviation of asymmetry from uniformity:
\beq
E = E' + \frac{\alpha}{ \rho_0} \int d{\bf r} \left( \delta
\rho_a \right)^2
- \frac{1}{2} \int d{\bf r} \, \Phi (r) \, \delta \rho_a \, .
\label{pol}
\eeq
Here,
$\Phi$ is the Coulomb potential,
$\rho_a$ is the neutron-proton density difference,
$\rho_a = \rho_n - \rho_p$, and we exploit the relation
for the density changes
$\delta \rho_p = -\delta \rho_a/2$ following from the requirement
$\rho_n +
\rho_p = \rho_0$.  The energy $E'$ is the remaining portion of
the net energy minimized with respect to the global partition of
asymmetry between the interior and surface.
The minimization of (\ref{pol}) with respect to $\delta
\rho_a$,
under the constraint of $\int d{\bf r} \, \delta \rho_a = 0$,
yields
\beq
\delta \rho_a (r) = \frac{\rho_0}{4 \alpha} \left(\Phi(r)
- \overline{ \Phi } \right)
= \rho_0 \, \frac{a_C }{8 \alpha } \frac{ Z}{ A^{1/3} }
\left(1 - \frac{5}{3} \frac{r^2}{R^2} \right) \, ,
\label{drho}
\eeq
where the overline indicates an average over the interior
volume and where we utilize the sharp-sphere potential $\Phi$.
On calculating contribution of the deviation
of asymmetry (\ref{drho}) from
uniformity, to the
difference of rms neutron and proton radii, we find
\beq
\frac{\delta \left(\langle r^2 \rangle_n^{1/2} -
\langle r^2 \rangle_p^{1/2} \right)}
{\langle r^2 \rangle^{1/2}}
= - \frac{a_C}{168\alpha} \frac{A^{5/3}}{N} \, .
\label{2Cou}
\eeq
Notably, the contribution from the polarization (\ref{drho})
to the net
energy is practically ignorable.

On combining (\ref{rnp}) with (\ref{2Cou}) and on incorporating
a correction for the surface diffuseness into (\ref{rhor}), we
get for the difference of the neutron and proton rms radii:
\beq
\frac{\langle r^2 \rangle_n^{1/2} -
\langle r^2 \rangle_p^{1/2}}
{\langle r^2 \rangle^{1/2}}
= \frac{A}{6NZ} \, \frac{N-Z}{1 + \frac{\beta}{\alpha} \,
A^{1/3}}
- \frac{a_C}{168 \alpha} \, \frac{A^{5/3}}{N} \,
\frac{\frac{10}{
3} + \frac{\beta}{\alpha} \, A^{1/3}}{1 + \frac{\beta}{\alpha}
\, A^{1/3}}
+ \pi^2 \, \frac{d \, (d_n - d_p)}{\langle r^2 \rangle}
\, .
\label{rmsnp}
\eeq
Calculations \cite{oya98} appear to indicate that, for lighter
nuclei, the
difference in the neutron and proton diffuseness $d_n-d_p$ is
of the second order in the difference of sharp-sphere radii
(\ref{rnp}).  For heavy nuclei, there may be a Coulomb
contribution to the difference in the diffuseness.  Overall,
in any case, the correction for diffuseness is of a lower
order in $A$ than the leading term in the skin
thickness.

\begin{figure}
\centerline{\includegraphics[angle=0,
width=.70\linewidth]{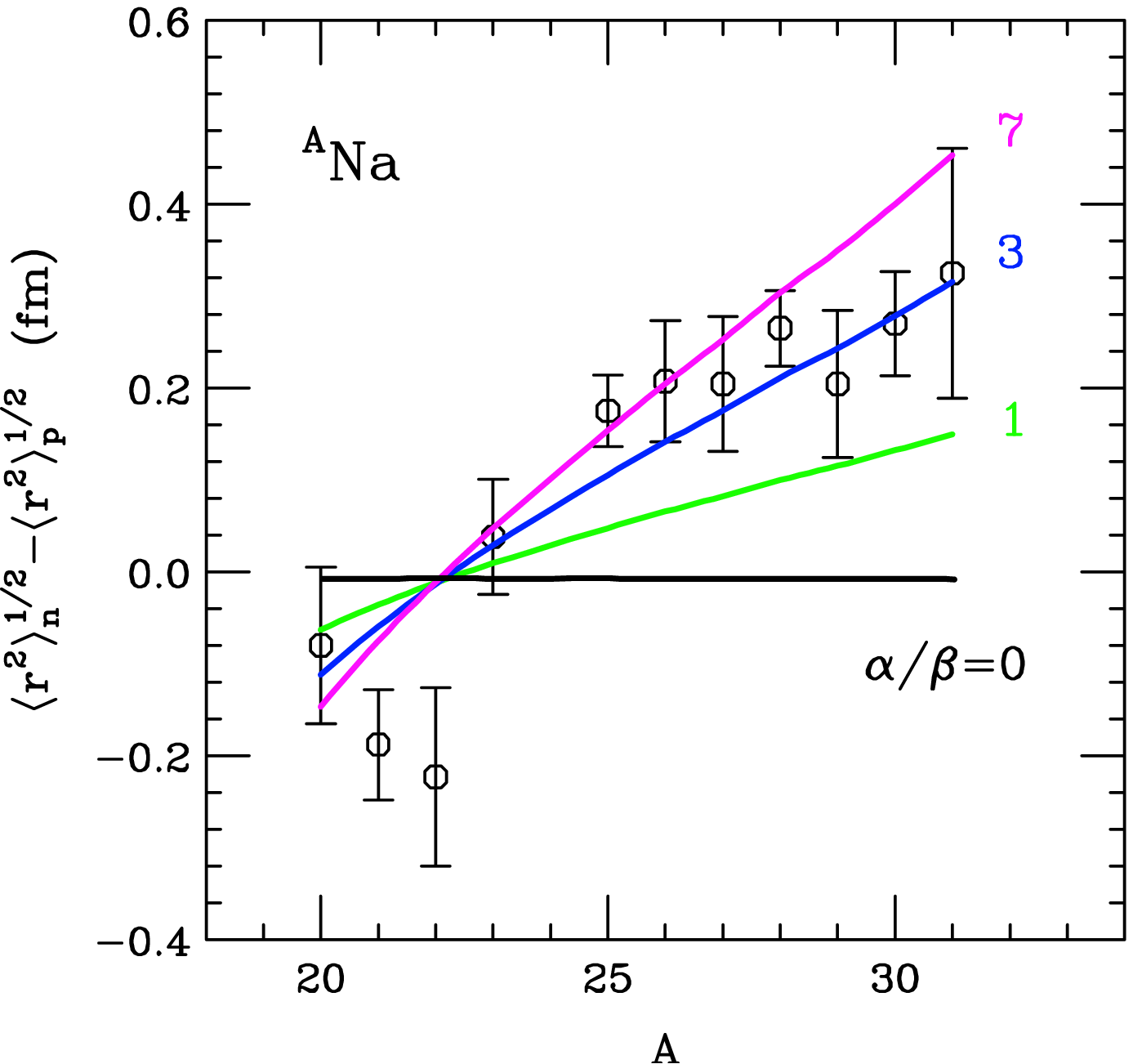}}
\vspace*{.2in}
\caption{
Difference between the neutron and proton rms radii for Na
isotopes as a function of the mass number.  The symbols
represent results from the data analysis Ref.\
\protect\cite{suz95}.  The different lines represent results
from Eq.\ (\protect\ref{rmsnp}) for the indicated values of
$\alpha/\beta$, $d_n-d_p=0$ and $\alpha$ taken from the
binding-energy correlation valley in Fig.\
\protect\ref{fig:rabc}.
}
\label{fig:narnp}
\end{figure}

\begin{figure}
\centerline{\includegraphics[angle=0,
width=.70\linewidth]{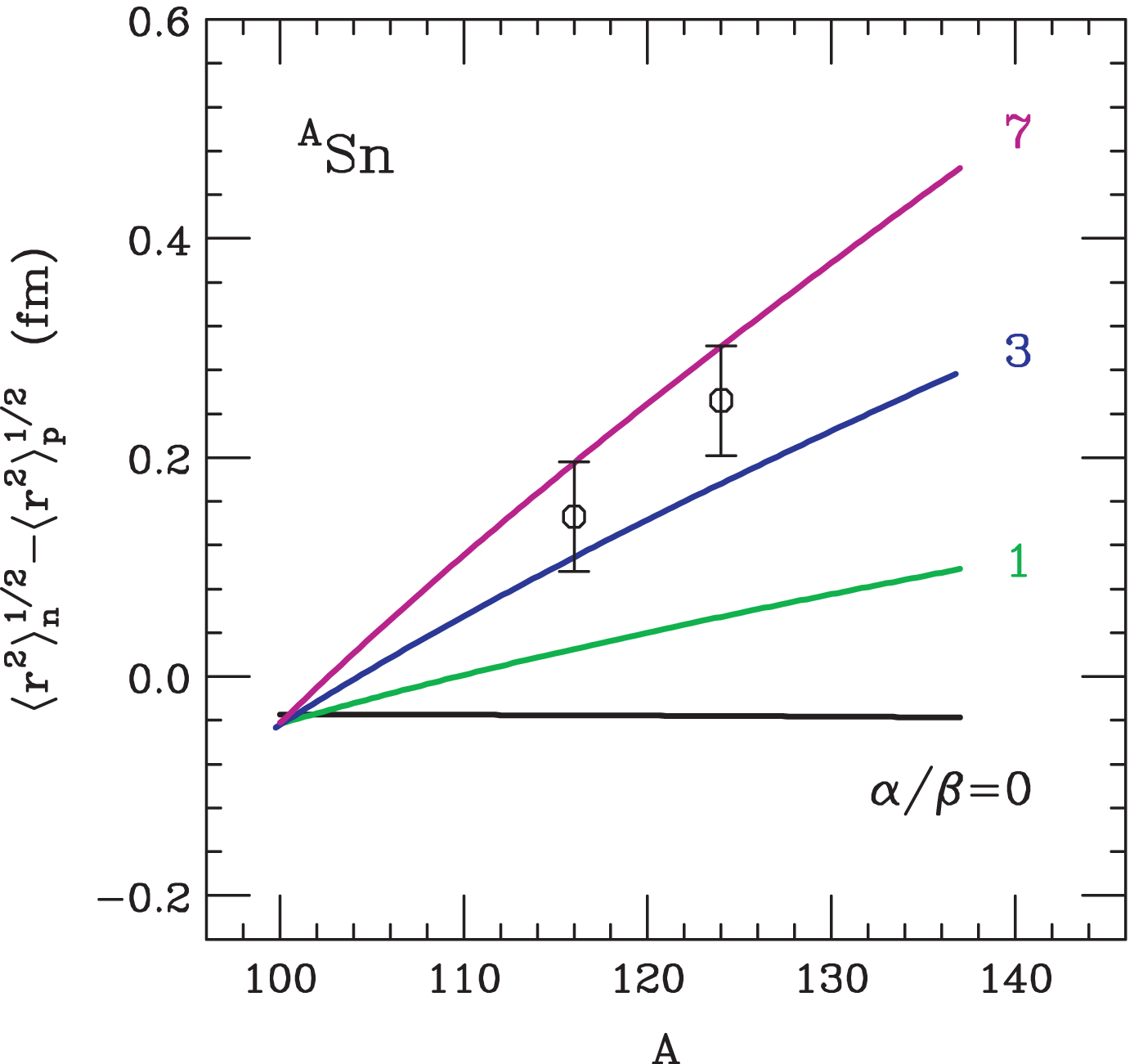}}
\vspace*{.2in}
\caption{
Difference between the neutron and proton rms radii for Sn
isotopes as a function of the mass number.  The symbols
represent results from the data analysis of Ref.\
\protect\cite{ray79}.  The different lines represent results
from Eq.\ (\protect\ref{rmsnp}) for the indicated values of
$\alpha/\beta$, $d_n-d_p=0$ and $\alpha$ taken from the
binding-energy correlation valley in Fig.\
\protect\ref{fig:rabc}.
}
\label{fig:snrnp}
\end{figure}

\begin{figure}
\centerline{\includegraphics[angle=0,
width=.70\linewidth]{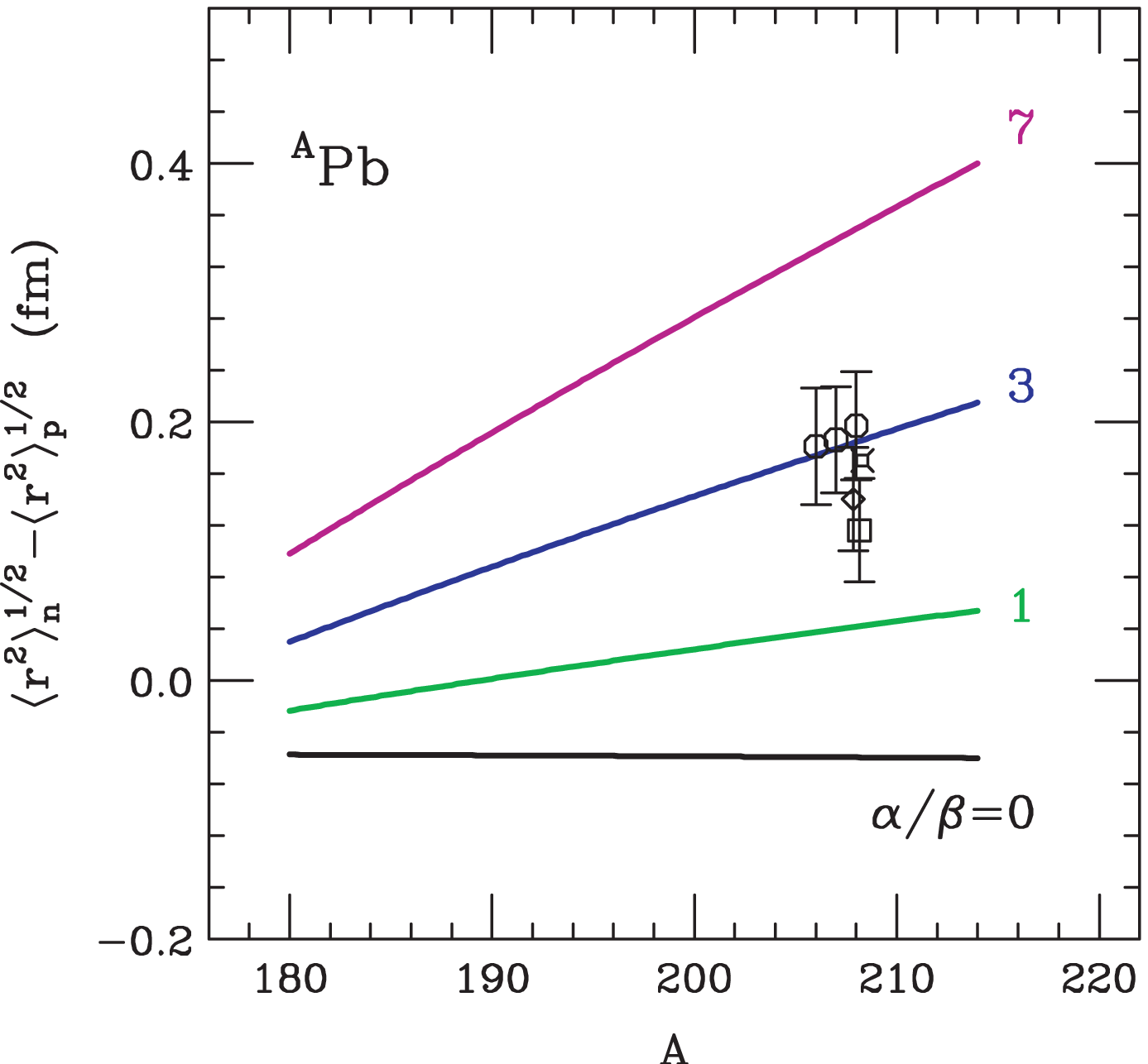}}
\vspace*{.2in}
\caption{
Difference between the neutron and proton rms radii for Pb
isotopes as a function of the mass number.  The symbols
represent results from the data analysis of Refs.\
\protect\cite{sta94} (circles), \protect\cite{ray79} (a diamond),
\protect\cite{cla02} (a square) and \protect\cite{kar02} (a cross).
The different lines represent results
from Eq.\ (\protect\ref{rmsnp}) for the indicated values of
$\alpha/\beta$, $d_n-d_p=0$ and $\alpha$ taken from the
binding-energy correlation valley in Fig.\
\protect\ref{fig:rabc}.
}
\label{fig:pbrnp}
\end{figure}

In Fig.\ \ref{fig:narnp}, we compare the results of skin
measurements in Na isotopes \cite{suz95} to the results from
(\ref{rmsnp}) for
different assumed values of $\alpha/\beta$, with the difference
in diffuseness set to zero and with $\alpha$
made to follow the binding-energy correlation valley in Fig.\
\ref{fig:rabc}.  The data rather clearly rule out $\alpha/\beta
< 1$, but it is difficult, so far, to determine the energy
parameter ratio with
a much better precision.  Further comparisons to data are shown
in Figs.\ \ref{fig:snrnp} and \ref{fig:pbrnp}, for Sn and Pb
isotopes, respectively.
Overall, the displayed data appear to
favor the ratios in the vicinity of $\alpha/\beta \sim 3$.
As the charge number increases, one can see in
Figs.~\ref{fig:narnp}-\ref{fig:pbrnp}, from the $\alpha/\beta=0$ results, the
increasing importance of Coulomb corrections to the
skin size.  For realistic ratios $\alpha/\beta \lesssim 3$, we find that
the Coulomb forces contribute about 0.07 fm of reduction to the
skin of $^{208}$Pb.  This can be compared to the $\sim 0.1$ fm
change in the skin size found by Furnstahl \cite{fur02} for
$^{208}$Pb when
switching off the Coulomb interactions within mean-field
models.  The interior polarization and pushing out of the
proton surface by the Coulomb interactions contribute about
evenly to the skin reduction in our estimates.

It needs to be mentioned that, when different
methods are employed in the skin extraction from measurements,
such as in the
case of Ca isotopes \cite{ray79,shl79,bat89,gib92,cla02}, the
outcomes can easily differ by more than the claimed extraction errors,
underscoring the difficulty in extractions.  Results from a
global fit of (\ref{rmsnp}) to the available skin sizes
obtained from data
\cite{ray79,shl79,bat89,gib92,sta94,suz95,kar02,cla02}
are next shown in terms of contour lines (solid, slanted downward) in Fig.\
\ref{fig:rabc}. As has been already discussed,
the dependence of skin sizes on asymmetry primarily constrains the
values of the ratio $\alpha/\beta$.  We provide the contours for two
combinations of a standard statistical deviation and an assumed
average theoretical error $\delta$ of (\ref{rmsnp}), i.e.\
$\chi^2 = \chi_{min}^2 + 1 + \sum_k (\delta/\sigma_k)^2$ where
$k$ is data index, and $\delta = 0.02$ or 0.04~fm.

\subsection{Skin Size vs Difference of Separation Energies}

A strong dependence of the sizes of asymmetry skins on the proton-neutron
difference of separation energies
was predicted within the mean-field calculations of Ref.\
\cite{tan92}.  The dependence has been observed
experimentally \cite{suz95} and it was referred to in other
mean-field calculations \cite{bro96,bro00}.
As we shall next demonstrate, this type of dependence
naturally emerges in our results.  The dependence can be used
to constrain directly the surface
parameter~$\beta$.

The proton-neutron separation energy difference, up to a factor and ${\mathcal
O}(A^{-1})$ effects, is nothing else but the asymmetry chemical
potential $\mu_a$.  From the energy contributions (\ref{ESnz})
and (\ref{ECVS}), we get right away
\beq
\mu_a = \frac{\partial E}{\partial (N_S - Z_S)} =
\frac{2 \beta \, (N_S - Z_S)}{A^{2/3}} -
\frac{5}{6} \, a_C \, \frac{Z}{A^{1/3}} \, .
\label{muaNZS}
\eeq
The difference of sharp-sphere radii (\ref{rnp}) can be then
written in terms of the chemical potential as
\beq
\frac{R_n - R_p}{R}
= \frac{A}{6NZ} \, (N_S-Z_S) =
\frac{\mu_a}{12 \beta} \,
\frac{A^{5/3}}{NZ}
+ \frac{5}{72} \, \frac{a_C}{\beta} \, \frac{A^{4/3}}{N} \, .
\eeq
Turning to the difference of the r.m.s.\ radii,
with the polarization contribution (\ref{2Cou}),
the diffuseness correction and with $\mu_a =
(S_p - S_n)/2$, where $S_p$ and $S_n$ are the proton and neutron
separation energies, we find
\beq
\frac{\langle r^2 \rangle_n^{1/2} -
\langle r^2 \rangle_p^{1/2}}
{\langle r^2 \rangle^{1/2}}
= \frac{S_p - S_n}{24 \beta} \, \frac{A^{5/3}}{NZ}
- \frac{a_C}{168 \alpha} \, \frac{A^{5/3}}{N} \,
\left(1 - \frac{35}{3} \, \frac{\alpha}{\beta \, A^{1/3}}
\right)
+ \pi^2 \, \frac{d \, (d_n - d_p)}{\langle r^2 \rangle}
\, .
\label{rmsnpmu}
\eeq

If not for the Coulomb correction, the surface asymmetry in (\ref{muaNZS}) would
have been proportional to the difference of separation energies, with the
coefficient of proportionality inversely proportional to $\beta$.  The linearity
with respect to the separation-energy difference is largely retained for the
skin size in (\ref{rmsnpmu}).  (Principally, one might strive in Eq.\ (\ref{rmsnpmu})
to eliminate the separate dependence on $N$ and $Z$, in favor of exclusively
$S_n - S_p$ and $A$, but the separate dependence on $N$ and $Z$ appears only in correction
terms, either associated with the Coulomb forces or geometry, so it is not crucial.)
Into the results previously compared to data, i.e.\
the dependence of energy and of skin size on mass and charge numbers,
the surface and volume symmetry coefficients entered in a combination.
By contrast, the dependence of skin size on the separation energy difference, that we now examine,
tests primarily the surface symmetry coefficient.  The difference between the cases
is easily understood following the capacitor analogy.  Thus, the volume
and surface coefficients characterize the surface and volume capacitances, respectively.  Either
the net energy or the ratio of the surface-to-net asymmetry depended on the net
capacitance.  However, in the capacitor analogy, the separation-energy difference
represents a voltage and the relation between surface charge and the voltage
depends only on the surface capacitance (as a matter of capacitance definition).

\begin{figure}
\centerline{\includegraphics[angle=0,
width=.71\linewidth]{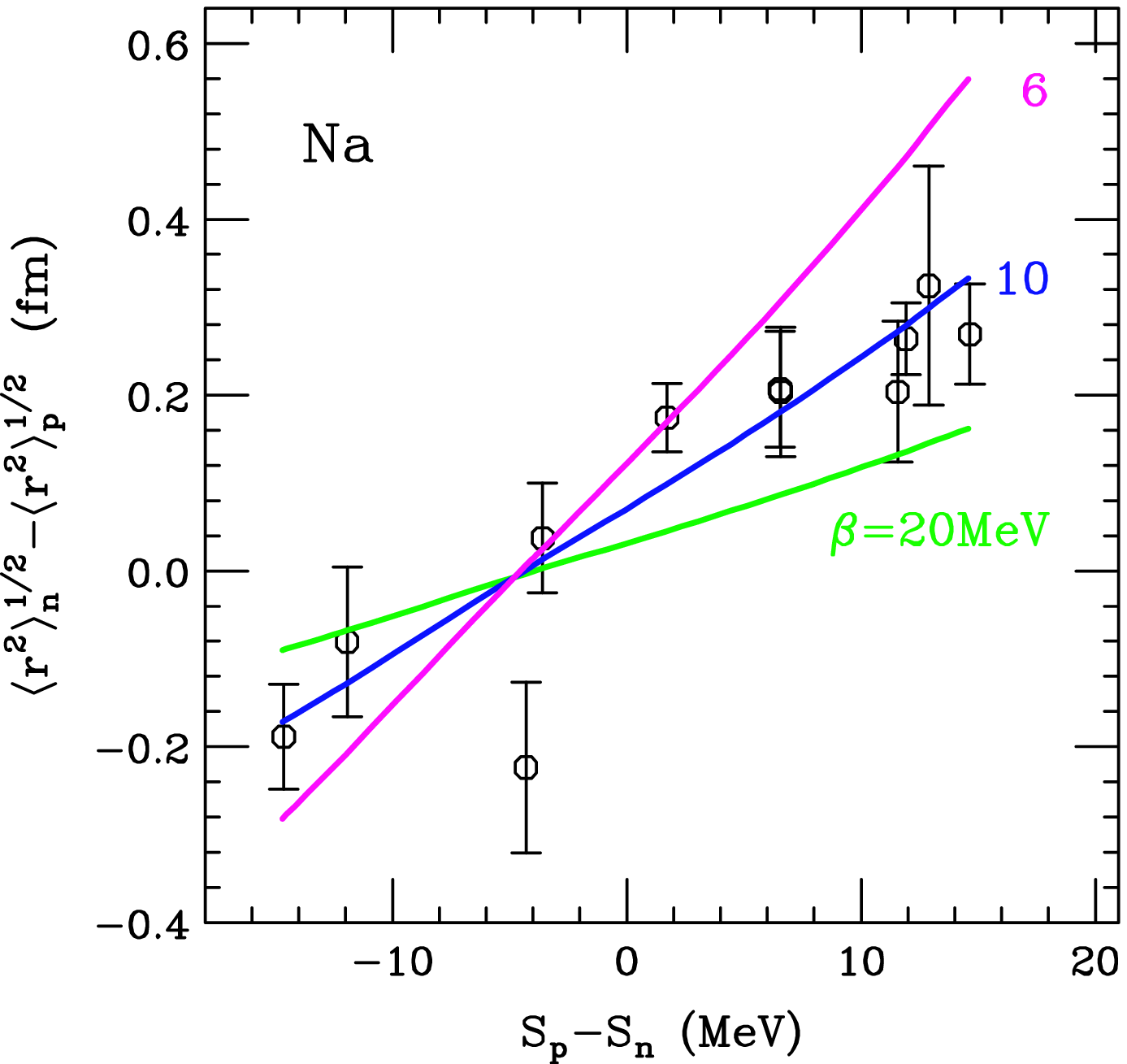}}
\vspace*{.2in}
\caption{
Difference between the neutron and proton rms radii vs the
difference between the proton and neutron separation energies
for Na isotopes.  The symbols represent data of Ref.\
\protect\cite{suz95}.  The lines represent predictions of Eq.\
(\protect\ref{rmsnpmu}) for different indicated values of
$\beta$ (in MeV) with the values of $\alpha$
made to follow the binding-energy correlation in Fig.\
\protect\ref{fig:rabc}.
}
\label{narm}
\end{figure}

\begin{figure}
\centerline{\includegraphics[angle=0,
width=.71\linewidth]{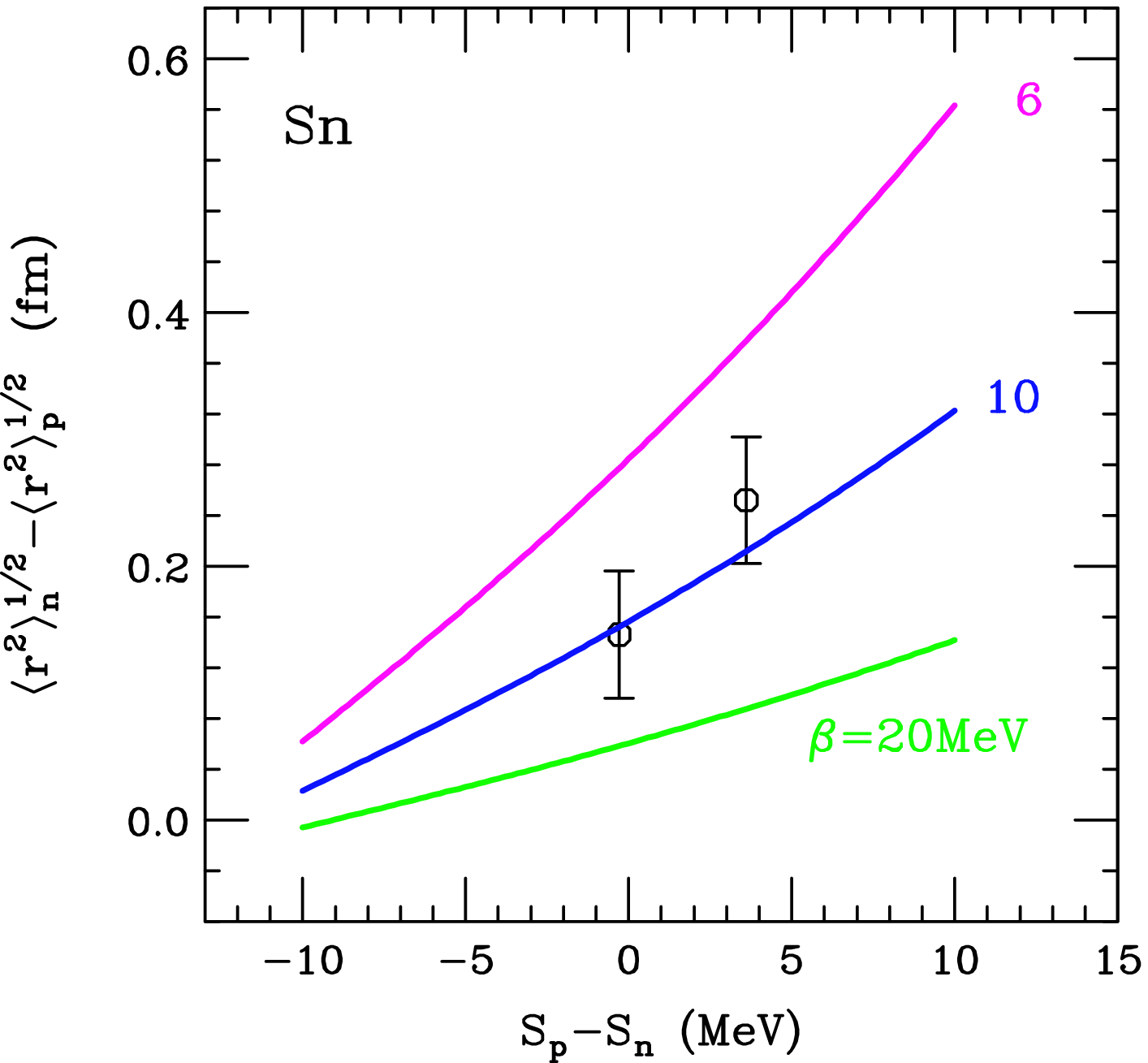}}
\vspace*{.2in}
\caption{
Difference between the neutron and proton rms radii vs the
difference between the proton and neutron separation energies
for Sn isotopes.  The symbols represent data of Ref.\
\protect\cite{ray79}.  The lines represent predictions of Eq.\
(\protect\ref{rmsnpmu}) for different indicated values of
$\beta$ (in MeV) with the values of $\alpha$
made to follow the binding-energy correlation in Fig.\
\protect\ref{fig:rabc}.
}
\label{snrm}
\end{figure}

\begin{figure}
\centerline{\includegraphics[angle=0,
width=.71\linewidth]{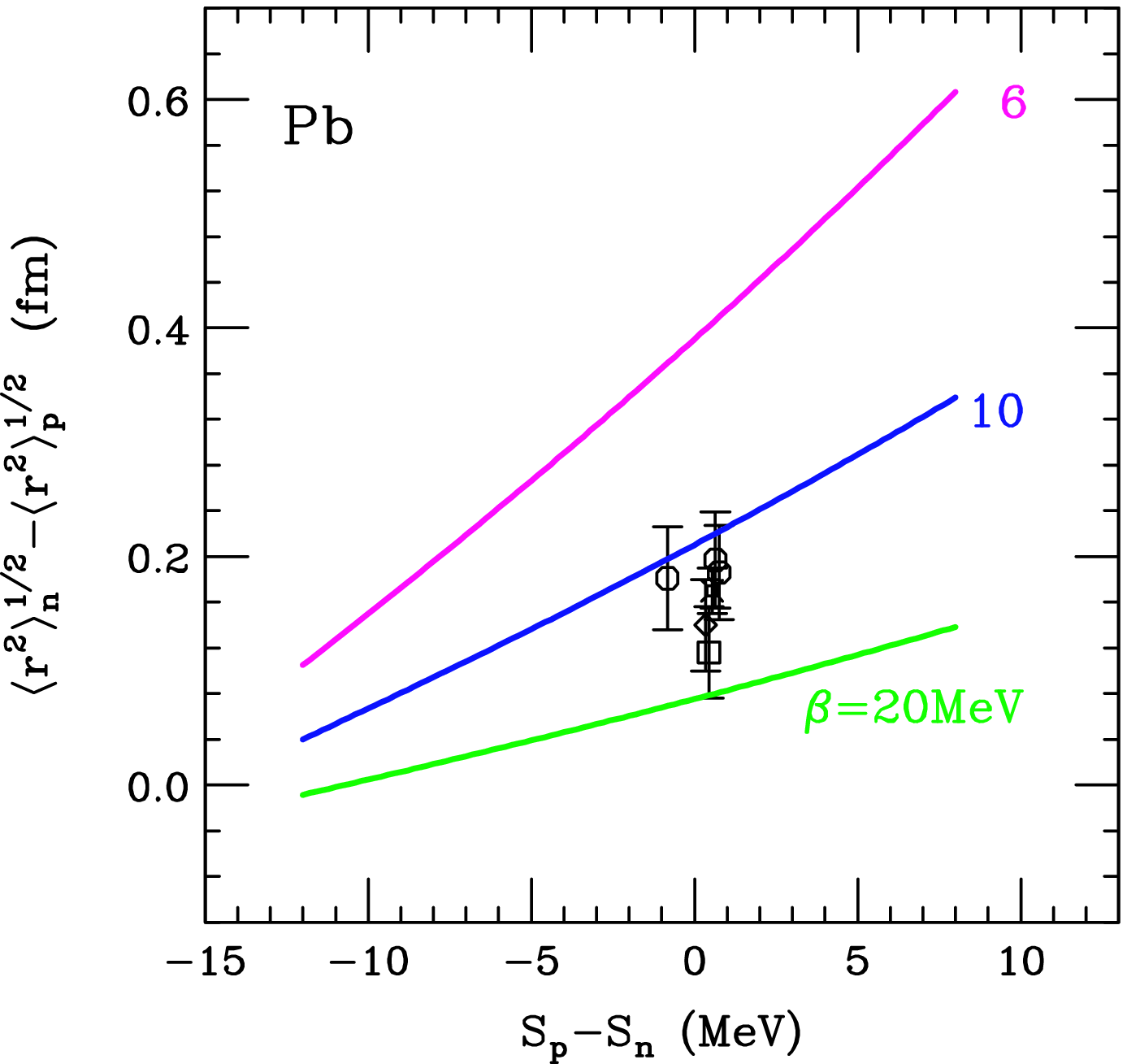}}
\vspace*{.2in}
\caption{
Difference between the neutron and proton rms radii vs the
difference between the proton and neutron separation energies
for Pb isotopes.
The symbols
represent results from the data analysis of Refs.\
\protect\cite{sta94} (circles), \protect\cite{ray79} (a diamond),
\protect\cite{cla02} (a square) and \protect\cite{kar02} (a cross).
  The lines represent predictions of Eq.\
(\protect\ref{rmsnpmu}) for different indicated values of
$\beta$ (in MeV) with the values of $\alpha$
made to follow the binding-energy correlation in Fig.\
\protect\ref{fig:rabc}.
}
\label{pbrm}
\end{figure}

We now confront the result (\ref{rmsnpmu}) with the data on skin sizes as a function
of separation-energy difference for Na, Sn and Pb nuclei in Figs.\ \ref{narm},
\ref{snrm} and \ref{pbrm}, respectively.  The data appear to favor the values $\beta \gtrsim
10$~MeV.  In fitting all data on skin sizes \cite{ray79,shl79,bat89,gib92,sta94,suz95,kar02,cla02},
now with Eq.\ (\ref{rmsnpmu}),
we again generate contour lines (solid, slanted
upward) in the plane of $\alpha$-$\alpha/\beta$ in Fig.\ \ref{fig:rabc},
for $\chi^2 = \chi_{min}^2 + 1 + \sum_k (\delta/\sigma_k)^2$ where
$k$ is data index, and the systematic error equal either to $\delta = 0.02$ or 0.04~fm.
Since the dependence primarily constraints $\beta$, the constraint lines
in the plane $\alpha$-$\alpha/\beta$ are approximately straight lines
passing through the origin of the coordinate set.

\subsection{Symmetry Parameter Values}

With the three dependencies fitted to the data, we now have
three regions in the plane of Fig.\ \ref{fig:rabc}, favored by the data,
that intersect.  Since the separation energies stem from the binding
energies, one would expect that the intersection of the two regions derived
from skin sizes would be well centered over the region favored by binding
energies.  Some offset may be from a bias due to the fact that the skin data
primarily stem from nuclei near closed shells or due to the
the fact that the surface diffuseness is ignored in the Coulomb corrections for
skin size but is not ignored in the binding-energy formula.  In any case, the offset
suggests that we may need to use bias errors as large as $\delta=0.03$~fm for
Eqs.\ (\ref{rmsnp}) and (\ref{rmsnpmu}).  Intersection of the three
regions favored by data then yields the following limits on the parameter values: $2.0 \lesssim
\alpha/\beta \lesssim 2.8$ for the ratio, $27 \, \mbox{MeV} \lesssim \alpha \lesssim 31 \,
\mbox{MeV}$ for the volume and
$11 \, \mbox{MeV} \lesssim \beta \lesssim 14 \, \mbox{MeV}$ for the surface.
Within those limits, though, the parameters are correlated, cf.\ Fig.\ \ref{fig:rabc}.

It should be mentioned that values of the volume symmetry parameters $\alpha$
in some advanced mass descriptions fall within our region, such as in
\cite{gor02} with $J=28\, \mbox{MeV}$ and in \cite{pom02} with their $-\kappa_{vol} \,
b_{vol} = 29.3 \, \mbox{MeV}$.  However, in \cite{gor02} the particular
parameter value is set and in
neither of those sources the uncertainties are actually assessed.  Elsewhere,
values outside of
our range are followed \cite{mol95} and in the mean-field structure
calculations, based on Skyrme interactions or
Lagrangians, a wide range of the symmetry volume parameters
is employed~\cite{fur02}, from 24.5 to 43.5~MeV.  Wide parameter ranges are
also employed in various phenomenological considerations such as for
the nuclear liquid-gas phase-transition \cite{mul95}.
On the other hand, nuclear-matter calculations relying on realistic nucleon-nucleon interactions
appear to yield volume symmetry energy values within our range, of 27-29~MeV
depending on the interaction
in \cite{eng97} and $30.10\,\mbox{MeV}$ in \cite{akm98}.  Those microscopic energies, however, are
obtained as the difference between the neutron and symmetric nuclear-matter energies but this,
normally, is an excellent approximation to the symmetry energy from $N-Z$ expansion.

Regarding the obtained values of the ratio of $\alpha/\beta$,
one may ask about a justification
for the large-$A$ expansion of the symmetry energy in mass
formulas \cite{mol95,pom02}.
With the expansion
amounting to an approximation in (\ref{weiz2}) of $1/(1+x) =
1-x + x^2 - \ldots \approx 1-x$, where $x=\alpha/(\beta\, A^{1/3})$, we may
adopt a criterion that the expansion is justified when the
correction to the effect of surface on symmetry energy,
from ignoring higher-order terms, is less than
20\%.  This requires $x < 0.2$ or $A> (5 \alpha/\beta)^3 \ge 1000$
which is never satisfied.  (Even when the next term in the expansion is
included \cite{pom02}, meeting the criterion necessitates
$A> 5^{3/2} \, (\alpha/\beta)^3 =$ (90-250), depending on the $\alpha/\beta$
ratio.)
In the droplet
model fit \cite{mol95} to binding energies and
fission barriers, with nonexpanded symmetry energy,
the value of the ratio $\alpha/\beta$ is within
our region, as represented by $\frac{9}{4} \, \frac{J}{Q} = 2.5$, and so is the
value of the surface parameter $\beta$, represented by $\frac{4}{9} \, Q= 13.0 \,
\mbox{MeV}$.  However, the uncertainties are not produced there
and, moreover, with an inclusion of the volume symmetry parameter, $J=32.7\,
\mbox{MeV}$, the results are well
outside of our correlation region.  In earlier mass and barrier fits
within the droplet model, the ratio itself and the surface parameter
were much different, with $
\frac{9}{4} \, \frac{J}{Q} = 4.9$ and $\frac{4}{9} \, Q= 7.5 \,
\mbox{MeV}$ in \cite{mye75}, respectively.
It would be interesting to find out the value of the surface symmetry
coefficient for the advanced mass calculations in \cite{gor02}.  In \cite{pom02},
the volume-to-surface symmetry parameter ratios $\alpha/\beta$ for mass fits are
$-b_{surf} \, \kappa_{surf}/(b_{vol} \, \kappa_{vol})= 1.31$ and 1.35, but these
results are for an expanded form of the symmetry energy, at two different
levels of the expansion.

In the next section, we employ our elementary results in
the description of asymmetry oscillations in nuclei.

\section{Application to Asymmetry Oscillations}
\label{sec:reso}

\subsection{Goldhaber-Teller and Steinwedel-Jensen Models of
the Giant Dipole Resonance}

So far, we discussed elementary static properties of
ground-state nuclei.  Nuclear excitations may involve
displacements of protons relative
to neutrons, compared to the ground state, with prominent giant
resonance structures in cross
sections allowing for an interpretation in terms of a
collective motion involving
a number of nucleons.  The lowest in energy (or
frequency)
is the giant dipole resonance (GDR) prominent in the low-energy
photoabsorption \cite{die88} and inelastic electron-scattering
\cite{pit79} cross-sections.

\begin{figure}
\centerline{\includegraphics[angle=0,
width=.75\linewidth]{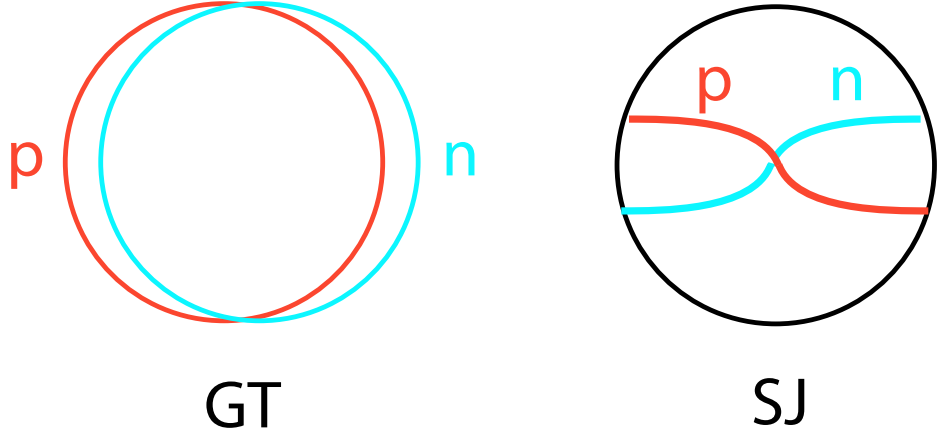}}
\vspace*{.2in}
\caption{
In the GT model \protect\cite{gol48} of the GDR, neutron
and proton distributions are
assumed to oscillate, unchanged with respect to their centers,
against each other.  In the SJ
model \protect\cite{gol48,ste50} of the GDR, particle
asymmetry oscillates within the nuclear
interior, with the flux of asymmetry vanishing at the
nuclear surface.
}
\label{fig:GDR}
\end{figure}

Two early models have been advanced for the GDR, under the names of
Goldhaber
and Teller (GT) \cite{gol48} and Steinwedel and Jensen (SJ)
\cite{gol48,ste50}.  In the GT model, the neutron and proton
distributions are assumed to oscillate against each other as
rigid entities, see Fig.~\ref{fig:GDR}.  Since the potential energy
associated with the displacement is proportional to the nuclear
surface area, i.e.\ $\propto A^{2/3}$, while the moving mass is
proportional to $A$, the resonance excitation energy in this
model is proportional to $A^{-1/6}$: $E_{GDR} = \hbar \Omega
\propto \sqrt{A^{2/3}/A} = A^{-1/6}$.  In the SJ model, a
standing wave of neutron vs proton displacement develops within
the nuclear volume, satisfying the surface condition of
vanishing
asymmetry-flux, analogous to the closed-pipe condition in the
standard example of oscillations for a continuous medium.  As
the
wavelength is then in a definite  constant ratio to the nuclear
radius, the resonance energy in the model is proportional to
$A^{-1/3}$: $E_{GDR} = \hbar c_a /\lambda \propto A^{-1/3}$,
where $c_a$ is the speed of propagation for asymmetry
perturbations in normal symmetric matter.

In detail, within the SJ model \cite{ste50,rin80}, it is
assumed that there is no {\em net} movement of matter, i.e.\ the net
matter velocity ${\bf v}$ vanishes and the net nucleon density
stays intact, consistent with the continuity equation:
$\partial \rho/\partial t = - {\bf \nabla} (\rho \, {\bf v})$.
The relative density $\rho_a = \rho_n - \rho_p$ changes, but
the state of matter is locally the same as within a ground-state
nucleus, just at the new asymmetry.
The vanishing net velocity implies a relation between the
velocities for individual particles, ${\bf v}_n$ and ${\bf
v}_p$, and densities, following from
the consideration of fluxes that need to meet the condition
\beq
0 = \rho \, {\bf v} = \rho_n \, {\bf v}_n + \rho_p \, {\bf v}_p
\, .
\eeq
This then implies that the flux in the asymmetry continuity
equation may be expressed in terms of the relative velocity
${\bf v}_{np} = {\bf v}_n - {\bf
v}_p$,
\beq
\frac{\partial \rho_a}{\partial t} = - {\bf \nabla}(\rho_n \,
{\bf v}_n - \rho_p \, {\bf v}_p)
= - {\bf \nabla} \left( \frac{2  \rho_n \, \rho_p}{\rho} \,
{\bf v}_{np} \right) \simeq - \frac{2 N  Z \rho_0}{A^2} \, {\bf
\nabla} {\bf v}_{np} \, .
\label{continuity}
\eeq
where the approximate equality is valid in the lowest order in
the disturbance.  An Euler type equation for the relative
velocity can be next used to close the equation above,
\beq
\frac{\partial}{\partial t} \,  {\bf v}_{np} = - \frac{2}{m^*}
{\bf \nabla} \mu_a = - \frac{2}{m^*} \, \frac{\partial
\mu_a}{\partial \rho_a} {\bf \nabla} \rho_a
= - \frac{2 \, c_a^2}{\rho_0} \, {\bf \nabla} \rho_a
\, ,
\label{Euler}
\eeq
where $m^*$ is the effective mass and
\beq
c_a^2 = \frac{\rho_0}{m^*} \, \frac{\partial \mu_a}{\partial
\rho_a} \, .
\label{c2}
\eeq
On differentiating Eq.\ (\ref{continuity}) with respect to time
and on substituting
(\ref{Euler}), we get
the wave equation for the asymmetry density
\beq
\frac{\partial^2}{\partial t^2} \, \rho_a = \frac{4NZ}{A^2} \,
c_a^2 \, \nabla^2 \rho_a \, .
\label{eq:wave}
\eeq

Under spherical symmetry, the normal vibrations are looked for,
in the standard way,
in terms of the deviation from ground-state density
expressed as the product of a spherical
Bessel function $j_\ell$ and of a spherical harmonic $Y_{\ell m}$,
\beq
\delta \rho_a = D_V \sin{(\omega t)} \, j_\ell(qr) \,
Y_{\ell m} (\Omega) \, ,
\label{drhojl}
\eeq
where, following (\ref{eq:wave}), the wavenumber $q$ and
the frequency $\omega$ are related
by
\beq
\omega = \frac{2\sqrt{NZ}}{A} \, c_a \, q \, .
\label{dispersion}
\eeq
The radial velocity component, from (\ref{Euler}) and (\ref{drhojl}), is
\beq
v_{np}^r = \frac{2D_V}{\rho_0} \, \frac{A}{2\sqrt{NZ}} \, c_a
\cos{(\omega t)} j_\ell'(qr) \, Y_{\ell m}(\Omega) \, .
\label{vrjl}
\eeq
The condition that the normal component of flux vanishes at the
surface \cite{boh75}, $v_{np}^r(R,\Omega)=0$, yields the
wavenumber values
\beq
q_{\ell n} = \frac{j_{\ell n}'}{R} \, ,
\eeq
where $n$ orders the values and
$j_{\ell n}'$ represents the $n$'s zero of the spherical
Bessel function derivative, $j_\ell' (j_{\ell n}')=0$.
For the resonance energy, we get, given (\ref{dispersion}),
\beq
E_{\ell n}= \hbar \, \omega_{\ell n}
= \frac{2\sqrt{NZ}}{A} \hbar \, c_a \, q_{\ell n}
= \frac{2 (NZ)^{1/2}}{A^{4/3}}\, \frac{\hbar \, c_a \, j_{\ell
n}'}{r_0} \, ,
\eeq
where,
with (\ref{c2}) and~(\ref{EVnz}),
\beq
c_a = \sqrt{\frac{2\alpha}{m^*}} \, .
\eeq
The GDR is characterized by $\ell=1$ and $n=1$.  With $j_{11}'
\simeq 2.08$, the GDR energy in the SJ model is then, finally,
\beq
E_{GDR} = \frac{2 (NZ)^{1/2}}{A^{4/3}}
\left(\frac{2\alpha}{m^*} \right)^{1/2} \frac{2.08 \,
\hbar}{r_0} \approx
\left(\frac{2\alpha}{m^*} \right)^{1/2} \frac{2.08 \,
\hbar}{r_0 \, A^{1/3}} \, .
\label{EGDRSJ}
\eeq

\begin{figure}
\centerline{\includegraphics[angle=0,
width=.71\linewidth]{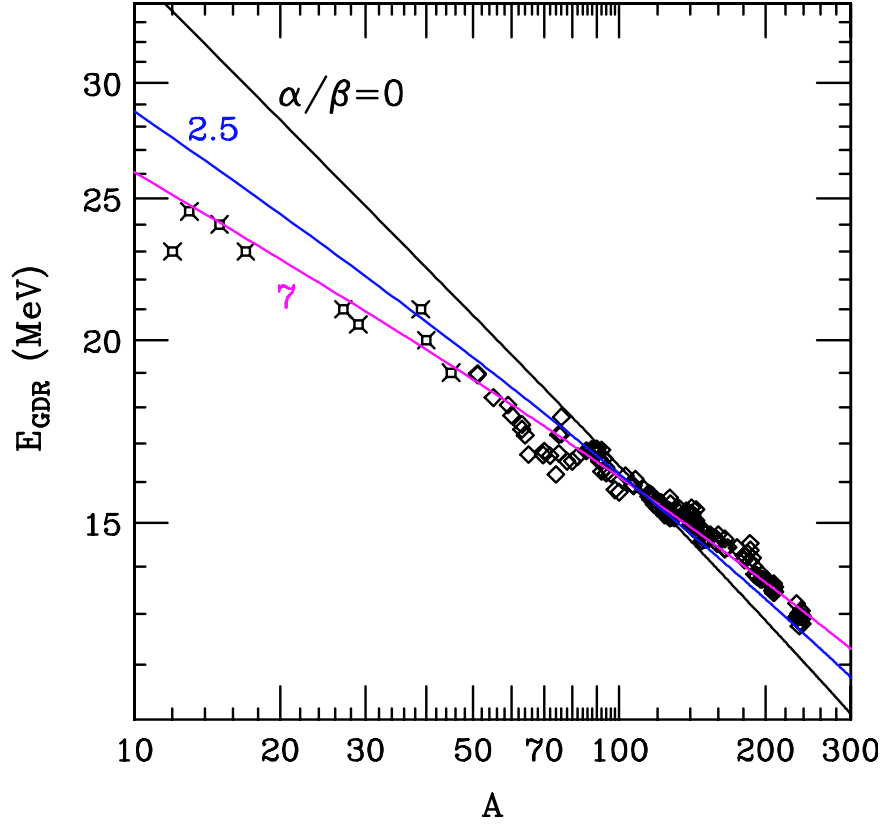}}
\vspace*{.2in}
\caption{
Energy of giant dipole resonance as a function of mass
number $A$.  Symbols represent photoabsorption data from the
compilation
\protect\cite{die88}.  For mass numbers $A > 50$, we use the
energies from the Lorentzian function fits from
\protect\cite{die88}.  When a splitting characteristic for
a deformed nucleus is present, we plot the equivalent energy for a
spherical nucleus.  For mass numbers $A < 50$, we plot the peak
values for nuclei for which a clear isolated peak is
present in the photoabsorption cross section
\protect\cite{die88}.  The lines represent energies
calculated from $E_{GDR}= (2\sqrt{(NZ)}/A) \, \hbar \, c_a
\, q_{11}$
with $c_a$ being a fit parameter and with $q_{11}$ obtained
from (\protect\ref{wavenumber}) at the different indicated values of
$\alpha / \beta$.
}
\label{fig:EGDR}
\end{figure}

Within the GT model of the GDR, the changes in asymmetry
occur
only within the nuclear surface and in the SJ model only within
the interior.
Experimental dependence of GDR energy on mass number, displayed
in Fig.\ \ref{fig:EGDR}, is
intermediate \cite{die88} between that characteristic for the
GT and SJ models, suggesting that both the volume and
surface participate in the oscillations.  To explain the mass
dependence of the GDR energy, in Ref.~\cite{mye77} the
oscillation in the SJ model was treated as an elementary mode
of nuclear oscillation coupled to a second GT mode of
oscillation, within
the same nucleus.  As the resonance frequencies for the two
models, GT and SJ,
shifted with a changing nuclear mass, the lower-frequency
normal mode for the combination of oscillators changed
from a mode with a predominantly GT to SJ content.
However, this casts the nuclear system into one with only
two degrees of freedom.  The physical situation is rather that of
a continuous system for which, though, the surface can generally accept
the asymmetry flux, making the original SJ boundary condition excessively
restrictive.

\subsection{Description of Asymmetry Oscillations}

We now explore the modification and consequences
of the surface boundary condition for the asymmetry
oscillations.
Given that the surface can store an asymmetry excess,
we
replace the SJ condition of a vanishing flux with
the condition for the change
in
surface asymmetry density
${\mathsf n}_a$,
\beq
\frac{\partial}{\partial t} {\mathsf n}_a =
\rho_n \,
{ v}_n^r - \rho_p \, { v}_p^r
=  \frac{2 N  Z \rho_0}{A^2} \,
 { v}_{np}^r \, ,
\label{nfluxR}
\eeq
cf.\ (\ref{continuity}).
As the chemical potential for the surface needs to be
the same as within the adjacent interior, this can be further
rewritten
as a condition involving the interior chemical
potential, or involving the volume density next to the surface,
\beq
\frac{2 N  Z \rho_0}{A^2} \,
 { v}_{np}^r
=
\frac{\partial}{\partial t} {\mathsf n}_a
=
\frac{\partial  {\mathsf n}_a}{\partial \mu_a} \,
\frac{\partial  \mu_a}{\partial t}
=
\frac{\partial  {\mathsf n}_a}{\partial \mu_a} \,
\frac{\partial  {\mu}_a}{\partial \rho_a} \,
\frac{\partial  \rho_a}{\partial t}
\, .
\label{fluxR}
\eeq
When the surface does not accept the excess, i.e.\
$\partial {\mathsf
n}_a /\partial \mu_a = 0$, we get the regular SJ condition
of a vanishing flux.

Looking for normal modes with (\ref{drhojl}) and
\beq
\delta {\mathsf n}_a = D_S \sin{(\omega t)} \,
Y_{\ell m} (\Omega) \, ,
\eeq
we find from the requirement of equality for chemical potentials at the surface,
in accordance with local surface and volume densities of symmetry energy,
that
\beq
D_S = \frac{\alpha}{3 \beta \, A^{1/3}} \, D_V \, R \, j_{\ell}
(qR) \,  .
\label{amplitudeSV}
\eeq
Moreover, with
(\ref{vrjl})
and (\ref{dispersion}), we can obtain from either
(\ref{nfluxR}) or
(\ref{fluxR}) the condition for the wavevector of oscillation:
\beq
qR \, j_\ell(qR) = \frac{3\beta \, A^{1/3}}{\alpha} \, j_\ell'
(qR) \, .
\label{wavenumber}
\eeq

As the ratio $\alpha/\beta A^{1/3}$ grows from small to large
values, the relative magnitude of the surface amplitude $D_S$
in (\ref{amplitudeSV}) grows.  Moreover, it is apparent that
the solutions to the wavenumber equation (\ref{wavenumber})
generally shift then
from $j_{\ell n}'/R$ to $j_{\ell n}/R$, where
$j_{\ell n}$
represents the $n$'th zero of the spherical
Bessel function, $j_\ell (j_{\ell
n})=0$.  In terms of the standard example of oscillations
for a continuous medium, the boundary condition shifts from one
representing a closed pipe to one representing an open pipe.
Coping with the $n=1$, $\ell>0$ solution to the wavenumber equation
at $q \rightarrow 0$, as $\beta A^{1/3}/\alpha \rightarrow 0$, however,
requires a careful expansion.

We concentrate next on the $\ell=1$ GDR case.
For $\alpha/\beta A^{1/3}
\rightarrow 0$, the surface amplitude in (\ref{amplitudeSV}) tends to zero and we arrive at
the SJ limit with the GDR energy given by (\ref{EGDRSJ}).  For
$\beta A^{1/3}/\alpha \rightarrow 0$, we need to consider the
Bessel function
in (\ref{wavenumber}) at small values of its argument where
$j_1(x) \approx (x/3) (1 -
x^2/10)$.  We then find for the wavenumber
\beq
q_{11} \simeq \left(\frac{3 \beta}{\alpha} \right)^{1/2}
\frac{1}{r_0 \, A^{1/6}} \, ,
\eeq
and for the resonance energy
\beq
E_{GDR} = \hbar \omega_{11} = \frac{2 \sqrt{NZ}}{A} \,
\hbar \, c_a \, q_{11} =
\frac{2(NZ)^{1/2}}{A^{7/6}} \,
\left(\frac{6 \beta}{m^*} \right)^{1/2} \frac{\hbar}{r_0}
\approx \left(\frac{6\beta}{m^*}\right)^{1/2}
\frac{\hbar}{r_0 \, A^{1/6}} \, .
\eeq
In this limit, the resonance energy becomes dependent only
on the surface symmetry parameter $\beta$ and not on the
volume parameter $\alpha$. Moreover, the volume amplitude
gets suppressed relative to the surface amplitude in
(\ref{amplitudeSV}).  This is clearly the GT limit of
vibration.

Figure \ref{fig:EGDR} shows next fits to the data \cite{die88}
on
mass dependence of the GDR energy, with $c_a$ treated as a fit
parameter and with $q_{11}$ obtained from (\ref{wavenumber})
for different values of $\alpha/\beta$.
We see that a realistic ratio of $\alpha/\beta=2.5$ yields a much
better description of the data than the SJ ratio
of $\alpha/\beta=0$.
However, a still larger ratio than 2.5 would provide an even better
description.
It is possible that other factors than the finite
capacitance of the surface for asymmetry play a role here, such as
the local deviations of a system from the ground state \cite{pin66}.
For the best fit with $\alpha/\beta=2.5$, we get $c_a=0.28\, c$,
corresponding to $m=0.84 \, m^*$.

One may ask for what values of $A$ and realistic $\alpha/\beta$
the vibrations turn from more GT to more SJ like.  As a
criterion, one may use the fraction of potential energy of the
vibration attributable to the surface.  With this criterion,
one finds that the transition takes place at $\beta
A_{tr}^{1/3}/\alpha = 3/2$, and, thus,
for the range of
$2.0 \lesssim \alpha/\beta \lesssim 2.8$, at $30 \lesssim A_{tr}
\lesssim 75$.  The relatively high transition masses can be attributed
to the node of the vibration at the
center of a nucleus.

\subsection{Transition Densities}

The profile of a standing wave in an oscillation represents, at
the quantal level, the transition density for the resonance.
This density is probed in inelastic electron scattering~\cite{pit79}.
To our knowledge, no inversion of data for the
GDR transition density has been ever made, but the data were tested
against the GT and SJ type densities, with a conclusion that the
GDR (and also $\ell > 1$) densities are in-between the two
limits \cite{pit79}.
Even in theoretical works, the transition densities have not
been much quoted.

We write the transition density as
\beq
\rho_1 ({\bf r}) = \rho_1 (r) \, Y_{\ell m} \, ,
\eeq
where $\rho_1 (r)$ includes both the volume and surface
contributions.  Interpreting the volume results in terms of
long-wavelength modulations of the particle densities and the
surface results in terms of the modulations of radii, we have
for the density \cite{pit79}
\beq
\rho_1 (r) = \frac{1}{\rho_0} \left[ D_V \, j_\ell (qr) \, \rho(r) - D_S
\, \frac{d \rho}{dr} \right] \,
\eeq
where $\rho(r)$ is the ground-state density distribution. With
(\ref{amplitudeSV}), we can further rewrite this as
\beq
\rho_1(r)= \frac{D_V}{\rho_0} \, j_{\ell}(qr)  \left[
\rho(r) - \frac{\alpha}{3 \beta \, A^{1/3}}\, r \,
\frac{d\rho}{dr} \right] \, ,
\label{rho1}
\eeq

\begin{figure}
\centerline{\includegraphics[angle=0,
width=.71\linewidth]{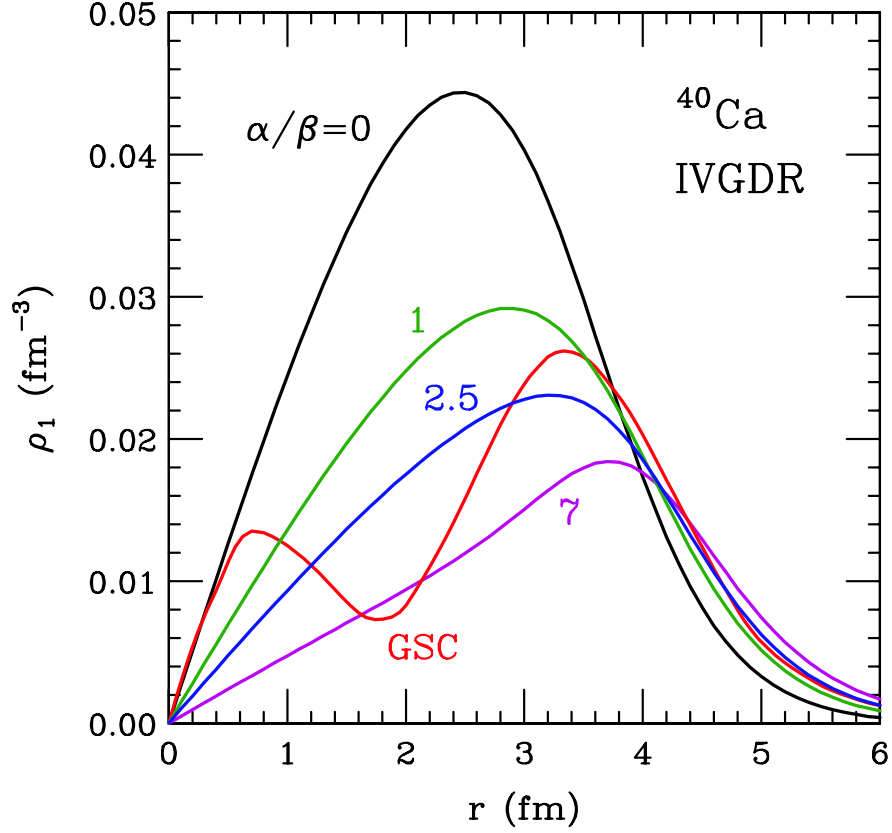}}
\vspace*{.2in}
\caption{
GDR transition density $\rho_1$ for $^{40}$Ca, as a function of
the distance $r$ from nuclear center.  The
lines marked with the respective values of parameter ratio
$\alpha/\beta$ represent the densities obtained from
Eq.\ (\protect\ref{rho1}).  The line marked GSC represents
density from the microscopic calculations
of Ref.\ \protect\cite{kam97},
including effects of the 2p-2h excitations and
ground-state correlations.
}
\label{fig:catd}
\end{figure}

Establishing a contact with microscopic theory, for $^{40}$Ca in
Fig.\ \ref{fig:catd}, we compare our GDR transition density (\ref{rho1}),
with (\ref{rhor}),
to the transition density
from the microscopic calculations of Ref.\ \cite{kam97}, that
include the effects of 2p-2h excitations and of
ground-state correlations.  The normalization in (\ref{rho1})
is obtained from the Thomas-Reiche-Kuhn sum rule; for illustrative
purposes, we take $m^*=m$ as in \cite{kam97}.
The calculations \cite{kam97} employ a
phenomenological Landau-Migdal type interaction in use in
describing different giant resonances.  We note that a best
contact with our results is made for $\alpha/\beta \gtrsim 2.5$.

\section{Density Dependence of Symmetry Energy}
\label{sec:micro}

Several works tried to identify links between the asymmetry
skins and bulk asymmetry properties of nuclear matter,
within microscopic theory
\cite{oya98,bro00,typ01,fur02}.  Having explored a link between
the skins and the symmetry parameters, we
will now try to establish a connection between the parameters
and the density dependence of symmetry energy, starting with
the Thomas-Fermi theory.

\subsection{Symmetry Parameters in the Thomas-Fermi Theory}
\label{ssec:symTF}
We shall consider a nuclear energy functional expanded to the
second order in asymmetry:
\beq
E = E_0 + \int d{\bf r} \, \rho \,
S(\rho)  \left( \frac{\rho_a}{\rho} \right)^2 \equiv E_0 + E_a \,
.
\label{EFT}
\eeq
Here we assume that the expansion term in
asymmetry is local, but this will not be essential for much of the
discussion that follows.  The function $S$ is the
density-dependent symmetry energy per nucleon and $S(\rho_0) =
\alpha$.  The energy $E_0$ is generally nonlocal but independent
of asymmetry as the zeroth-order term in the expansion; since the
Coulomb effects have been explicitly considered in Secs.\
\ref{sec:binding} and \ref{sec:skins}, we will disregard them for
the discussion that follows. Regarding the role of asymmetry in
any nonlocality, there had been indications \cite{oya98,fur02,bod03}
that it has a negligible impact on the properties of nuclei; the
apparent success in systematizing the asymmetry skins in terms of
the nuclear-matter properties \cite{oya98,bro00,typ01,fur02}
lends, in itself, a support to the weak impact.

A simple example of the functional (\ref{EFT}) is such as
employed in the initializations of transport calculations for
central reactions \cite{dan00}:
\bea
\label{E0model}
E_0 & = & \int d{\bf r} \, \left(\frac{3}{5} \, \epsilon_F \,
\rho + \frac{{\mathcal A} \, {\rho^2}}{\rho_0} + \frac
{{\mathcal B} \, \rho^{\sigma+1}}{(\sigma+1) \, \rho_0^\sigma}
\right) + \frac{{\mathcal D}}{2 \rho_0} \int d{\bf r} \, \left(
\nabla \rho \right)^2 \, ,  \\[.5ex]
S & = & \frac{1}{3} \, \epsilon_F + \frac{{\mathcal C} \, \rho}
{\rho_0} + \frac{{\mathcal F} \, \rho^2}{\rho_0^2} \, ,
\label{Smodel}
\eea
with the Fermi energy $\epsilon_F = (3\pi^2/2)^{2/3} \, \hbar^2
\, \rho^{2/3}/2m$.  The constants ${\mathcal A}$, ${\mathcal
B}$ and $\sigma$ are fixed by the requirement that the
functional yields the energy minimum of $E/A = a_V \approx
-16$~MeV at $\rho_0$ in symmetric nuclear matter,
characterized
by a specific incompressibility constant $K$.  E.g.\ for
$K=260$~MeV, we get ${\mathcal A} = -180.9$~MeV, ${\mathcal B}
= 128.1$~MeV and $\sigma=1.446$.  The constant ${\mathcal D}$
provides a size scale for the surface diffuseness and an adequate
surface profile is obtained \cite{dan00} for ${\mathcal D}
\approx 20.4$~MeV\,fm$^2$.  As to the constants in $S$, we have
the relation $\alpha = \epsilon_F(\rho_0) /3 + {\mathcal C} + {\mathcal F}$
and,
otherwise, different combinations of the constants ${\mathcal C}$
and
${\mathcal D}$ yield different values of $\beta$ as will be
discussed.

The minimization of the energy (\ref{EFT}) under the conditions of
fixed particle numbers
\beq
A = \int d{\bf r} \, \rho  \hspace*{2em} \mbox{and}
\hspace*{2em} N-Z = \int d{\bf r}\, \rho_a
\eeq
produces the Thomas-Fermi equations:
\bea
\label{TF}
\mu & = & \frac{\delta  E_0}{\delta  \rho} + \rho_a^2 \,
\frac{d}{d  \rho} \, \frac{S}{\rho}   \, ,  \\[.5ex]
\mu_a & = &
2 \, \rho_a \, \frac{S(\rho)}{\rho} \, .
\label{TFa}
\eea
At first, those equations will just serve as a background
illustration for the claims made in the discussion.

On the basis of (\ref{EFT}), we would like to find out changes in
the nuclear ground-state energy, as the particle asymmetry
changes.  First, we consider a symmetric system and find
$\rho({\bf r})$ that minimizes the energy $E = E_0$.  With the
energy $E$ being quadratic in asymmetry, the net density
minimizing the energy functional at a finite asymmetry will differ
from the density for the symmetric system only to within a second
order in asymmetry.  With this, to within the second order in
asymmetry, the ground-state $E_a$ in (\ref{EFT}) may be calculated
with $\rho({\bf r})$ obtained for symmetric matter. Moreover,
since the symmetric-system $\rho ({\bf r})$ minimizes $E_0$, any
changes to it, of the second order in asymmetry, will produce only
quartic terms of change in $E_0$. In conclusion, if we are
interested in the change in the ground-state energy to the second
order in asymmetry, we just need to consider the $E_a$ term in
(\ref{EFT}) and we can use there the density $\rho({\bf r})$ from
the energy $E_0$ minimized for a symmetric system.  In terms of
the Thomas-Fermi equations, this means that we can uncouple
(\ref{TF}) by putting $\rho_a =0$ there.

From the second of the Thomas-Fermi equations, we have in the
ground-state system
\beq
\rho_a = \frac{\mu_a}{2} \, \frac{\rho}{S} \,
\eeq
and from (\ref{EFT})
\beq
E_a = \frac{\mu_a^2}{4} \int d{\bf r} \, \frac{\rho}{S} \, .
\eeq
Within the analogy with a capacitor, the integral $\int d{\bf
r} \, (\rho/2S)$ represents the system capacitance, independent
of the deposited charge.  The volume capacitance is $A/2
\alpha$ and the reminder is the surface capacitance:
\beq
\frac{A^{2/3}}{2  \beta} \equiv \frac{{\mathcal S}}
{2 \pi \, r_0^2 \, \beta}
=
\frac{1}{2} \int d{\bf r} \, \rho \, \left( \frac{1}{S} -
\frac{1}{\alpha} \right) \, .
\label{Scap}
\eeq
As we will not be
pursuing here curvature effects, we just turn to the surface
capacitance for a semiinfinite matter.  With $x$ a coordinate
along the direction perpendicular to the surface and on factoring
out the density dependence from the symmetry energy with $S =
\alpha \, s(\rho/\rho_0)$, we find from (\ref{Scap})
\beq
\frac{\alpha}{\beta} = \frac{3}{r_0} \int dx \,
\frac{\rho}{\rho_0} \, \left( \frac{1}{s} - 1 \right) \, .
\label{abs}
\eeq
Thus, for (\ref{EFT}), given that the density
profile for the surface is constrained, the ratio of the symmetry
parameters $\alpha/\beta$ is a measure for the density dependence
of the symmetry energy~$S$. We note that a density-independent
symmetry energy, characterized by $s \equiv 1$, yields
$\alpha/\beta=0$.
The falling density at the surface produces in
this case no extra capacitance per nucleon compared to the
interior. The faster the drop of $s$ with density, the larger the
extra capacitance, exhibited in a larger $\alpha/\beta$ ratio.

\begin{figure}
\centerline{\includegraphics[angle=0,
width=.70\linewidth]{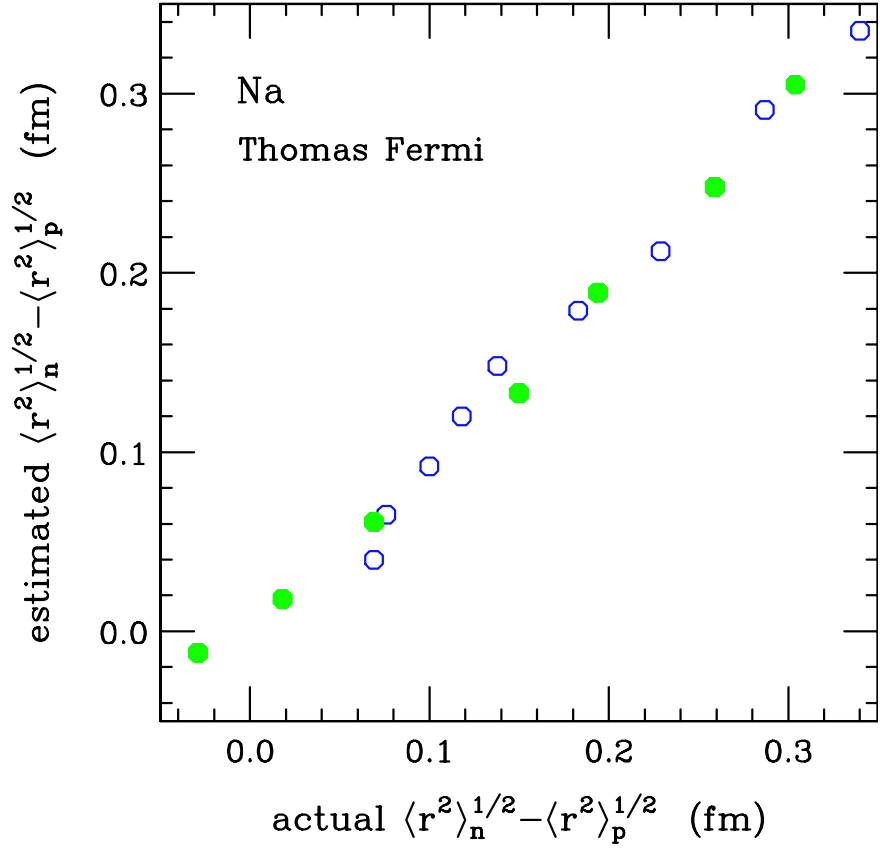}}
\vspace*{.2in}
\caption{
Size of the asymmetry skin estimated with Eq.\
(\protect\ref{rmsnp}) vs size obtained directly from the
calculated nucleon distributions, within the Thomas-Fermi calculations
of sodium isotopes.  The filled circles represent the results for
$\alpha/\beta=3$, while changing the isotope mass.  The open
circles represent the results for $^{30}$Na, when changing the values of
$\alpha/\beta$ and correlating the changes in $\alpha$ to follow
the valley in Fig.\ \protect\ref{fig:rabc}.
}
\label{fig:rr}
\end{figure}

We next test the validity of the considerations from Secs.\
\ref{sec:binding} and \ref{sec:skins}
within a Thomas Fermi model based on Eqs.\ (\ref{E0model}) and
(\ref{Smodel}), with a minor alteration, however, that the
different neutron and proton
Fermi energies are treated explicitly for asymmetric nuclei.  On
one hand, we solve the differential equation \cite{dan00} from
(\ref{TF}) for the semiinfinite symmetric matter and  we vary
${\mathcal C}$ and ${\mathcal F}$ to get various
combinations of $\alpha$ and $\alpha/\beta$ from (\ref{abs}),
requiring that they follow the binding-energy
correlation from Fig.\ \ref{fig:rabc}.  Given the $\alpha$ and
$\beta$ values, we make a prediction
for the size of the asymmetry skin of a specific nucleus,
following
Eq.\ (\ref{rmsnp}) with
$d_n-d_p=0$.  On
the other hand, we solve the Thomas-Fermi equations directly
for a nucleus, with Coulomb forces included, and calculate the
difference of rms radii.  A comparison of the results for
Na isotopes is shown in Fig.\ \ref{fig:rr}.
The results are both compared when changing the isotope mass, while
keeping the symmetry parameters fixed, and for one isotope when
changing the symmetry parameters.  As is seen, the agreement of
the estimate (\ref{rmsnp}) with the actual skin size is rather
remarkable.  For heavy nuclei, however, a small systematic
difference develops between the actual and estimated skin
sizes, that largely vanishes when the Coulomb forces are
switched off.  The size of the discrepancy, $\sim 0.03$~fm for
Pb, for which we have no understanding, is compatible with the
possible difference between our estimate of the Coulomb effects
in Eq.\ (\ref{rmsnp}) and that of Furnstahl \cite{fur02}
(already brought up in Sec.\ \ref{sec:skins}).

A very recent, more involved, analysis of surface symmetry energy and skin
sizes, within the Thomas-Fermi theory, has been carried out by Bodmer and
Usmani in Ref.\ \cite{bod03}, with some conclusions comparable to those that
emerge here.
In detailed comparisons of their results to ours, it should
be noted that their surface symmetry energy and parameter are differently
defined than ours for a finite nucleus and e.g.\ include
Coulomb contributions.

\subsection{Constraints on Density Dependence of Symmetry Energy}

To gain a perspective on the recent microscopic results on skin
systematics in the literature
\cite{bro00,typ01,fur02}, we further confront
our results from the Thomas-Fermi model, to those obtained in
the variety of mean-field models explored there.
In Fig.\ \ref{fig:rrapn}, the skins
from the mean-field calculations for $^{208}$Pb \cite{fur02} (symbols) are
presented vs the reduction factor $s$ for the symmetry energy
per nucleon at $\rho_0/2$
(left panel), vs the scaled derivative of the
energy at normal density $p_0 = \rho_0^2
\, dS/d \rho_0|_{\rho_0}$ (center panel) and vs the volume
symmetry
parameter
$\alpha$ (right panel).  The skin dependence on $p_0$ and on
$\alpha$ was explored in \cite{fur02}.  A similar derivative to
$p_0$ was utilized in \cite{bro00,typ01}.
Moreover in Fig.\ \ref{fig:rrapn}, the results from the Thomas-Fermi
model are presented (lines) for the symmetry parameters changing along
the correlation valley in Fig.~\ref{fig:alphad}.
It is apparent in each panel of Fig.\ \ref{fig:rrapn}, that
the Thomas-Fermi
model produces general trends such
as
found in the mean-field calculations.
However, in the three panels, the represented dependencies for
skins in the Thomas-Fermi
model are of a quite different nature.  Thus, when
$\alpha$ just provides a normalization
for the symmetry energy, it has a very limited bearing on the skins in
(\ref{rmsnp}), merely through a rescaling of the Coulomb
corrections (see also \cite{bod03}).  However, when $\alpha$ is correlated with
$\alpha/\beta$ through the binding-energy requirements, the
changes in $\alpha$ force changes in $\alpha/\beta$ and, thus,
produce significant changes in the skin in the rightmost panel of
Fig.\ \ref{fig:rrapn}.  In the center panel for the
Thomas-Fermi model, we observe combined indirect and direct
(albeit at the edge of the relevant density interval) effects
of the abscissa on the skin.  {Only} in the leftmost
panel for the Thomas-Fermi model, we observe the effect of a
direct physical connection between the abscissa and the skin,
through Eqs.\ (\ref{rmsnp}) and (\ref{abs}).  The three panels
illustrate potential dangers of scouting for a physical relation
\cite{bro00,typ01,fur02} looking at numerical results
within models that are necessarily constricted.

For a given effective interaction, the mean-field models likely yield
more realistic results in the surface region than the
Thomas-Fermi model.  However, there is no reason to believe
that a strong primary sensitivity
to the density dependence of $s$
disappears,
for the symmetry parameter
ratio $\alpha/\beta$,
in the mean-field models.  Using the results \cite{fur02} for the
$^{208}$Pb skins in those models, we can assess
$\alpha/\beta$ for the models, from (\ref{rmsnp}), and we can produce
constraints on the low-density $s$ from the constraints on
$\alpha/\beta$.  An alternative would be to use the data on
$^{208}$Pb skins directly to constrain $s$ in Fig.\
\ref{fig:rrapn}.  In Fig.\ \ref{fig:sab}, the values of $s$ at
$\rho_0/2$ are shown, for the mean-field
models,
vs
$\alpha/\beta$ from inverting (\ref{rmsnp}).  In the inversion, basing here
the results on one heavy nucleus, we augment
the Coulomb correction
in (\ref{rmsnp}) by 0.03~fm; an {\em ad hoc} procedure
suitable for all
nuclei might be to amplify the Coulomb correction by a factor
of~1.4.  Given the variation of the results in Fig.\
\ref{fig:sab}, for the acceptable range of $2.0 \lesssim
\alpha/\beta \lesssim 2.8$, we deduce the range of the symmetry-energy
reduction factor at $\rho_0/2$ of $0.57 \lesssim s \lesssim
0.83$.  Compared to the previous unbiased limit \cite{fur02} on $p_0$,
of $1\,\mbox{MeV/fm}^3 < p_0 < 4\, \mbox{MeV/fm}^3$, our limit on
$(1-s)$ represents a nearly twofold improvement in constraining
the low-density dependence of symmetry energy.  The $s$-shift is for
the fairness in comparison; note that the majority of theoretical
calculations in Fig.\ \ref{fig:sab} suggests an even narrower range
for $s$ than above, of $0.67 \lesssim s \lesssim 0.83$.
Concerning the microscopic nuclear-matter calculations with realistic
nucleon-nucleon interactions, minimally extrapolating underneath
the density range in \cite{eng97}, the reduction in symmetry
energy at $\rho_0/2$, for the variety of interactions there, is
within our narrow reduction range, at 0.67-0.73.  However, in
the calculations of Ref.\ \cite{akm98},
the reduction in symmetry energy at $\rho_0/2$ is only within our
wider range, at
0.59.

\section{Conclusions}
\label{sec:conc}

Binding in the symmetric nuclear matter can be realistically assessed by anyone
with a minimal experience, using the textbook mass formula
and a sample of nuclear masses.  We hope to have demonstrated
that the information on symmetry energy can be accessed with
a nearly comparable ease, using a minimally modified formula.
Elementary but general considerations,
for the nucleus as a macroscopic system,
 indicate that the dependence
of surface tension on nuclear asymmetry necessarily implies the
emergence of an asymmetry skin.  In the lowest order in asymmetry,
the surface tension and surface energy need to be quadratic in the asymmetry.
With analogous behavior of the volume energy, the surface and volume
symmetry energies combine as energies of two connected capacitors
with the volume capacitance proportional to the volume and the surface capacitance
proportional to the area.

In minimizing the net energy, the net asymmetry
partitions itself into the surface and volume
asymmetry in proportion to the capacitances.  As nuclear size
increases, the surface portion of the asymmetry decreases, as
surface decreases relative to the volume.  The minimal net symmetry energy is
given in terms of the net asymmetry squared divided by the sum of the
volume and surface capacitances.

In analyzing the binding energies, we demonstrate
that the surface symmetry contribution is required to describe
the energies of light asymmetric nuclei at a level similar to that typical for other nuclei.
Overall, however, the binding energies are better in constraining
a combination of the coefficients for the surface and volume symmetry
energies,
than in constraining the individual coefficients,
at least at the level of our
simple formula.  The specific coefficient combination
represents the symmetry coefficient in the standard mass formula.
On the other hand, the dependencies of skin sizes on asymmetry
and on the neutron-proton separation-energy difference primarily constrain, respectively, the
ratio of the symmetry volume to surface
parameters and the surface symmetry parameter.  Using the three
types of constraints, we establish limits on the volume symmetry parameter
of $27\, \mbox{MeV} \lesssim \alpha \lesssim 31 \, \mbox{MeV}$,
on the surface symmetry parameter of $11\, \mbox{MeV} \lesssim \beta \lesssim
14 \, \mbox{MeV}$, and on the parameter ratio, of $2.0 \lesssim \alpha/\beta \lesssim 2.8$.
Within those limits, however, the parameters are correlated.
In analyzing the skin sizes, we provide straightforward interpretations
for regularities observed both in measurements and in microscopic structure
calculations.

For reference, when taking the
parameter ratio in the middle of the permitted range, at
$\alpha/\beta \simeq 2.4$, the sample best-fit parameter values in
describing the binding energies with (\ref{weiz2}) and
(\ref{delta}) are $a_V=15.4808$~MeV, $a_S = 18.4359$~MeV,
$a_C=0.696424$~MeV, $a_P=9.82356$~MeV, $\alpha = 29.1048$~MeV and
$\beta=12.1270$~MeV.  When ignoring the surface diffuseness in
putting $\Delta=0$ in (\ref{weiz2}), the best-fit value are
$a_V=15.6163$~MeV, $a_S=17.9878$~MeV, $a_C=0.692791$~MeV,
$a_P=10.8714$~MeV, $\alpha=32.6655$~MeV and $\beta=13.6106$~MeV.

We point out that the additional capacitance for asymmetry provided by the nuclear
surface is associated with the drop of symmetry energy per nucleon with density.
The ratio of the volume to surface symmetry coefficients can be used to assess
the pace of that drop.  We test the analytic result for the skin size within the
Thomas-Fermi approximation and we use the results from microscopic mean-field
calculations to constrain the drop in the symmetry energy at half of the normal
density, $s=S(\rho_0/2)/S(\rho_0)$, finding $0.57 \lesssim s \lesssim 0.83$.

For the collective asymmetry oscillations, we derive a simple boundary condition,
following a consideration of the surface fed by the asymmetry flux from the
nuclear interior.  The nature of the condition depends on the nuclear mass, as
the relative capacitance of the surface for asymmetry changes with the system size.
As the size increases, the nature of the collective
oscillation changes from a surface to a volume mode.

\acknowledgements

\begin{figure}
\centerline{\includegraphics[angle=0,
width=.55\linewidth]{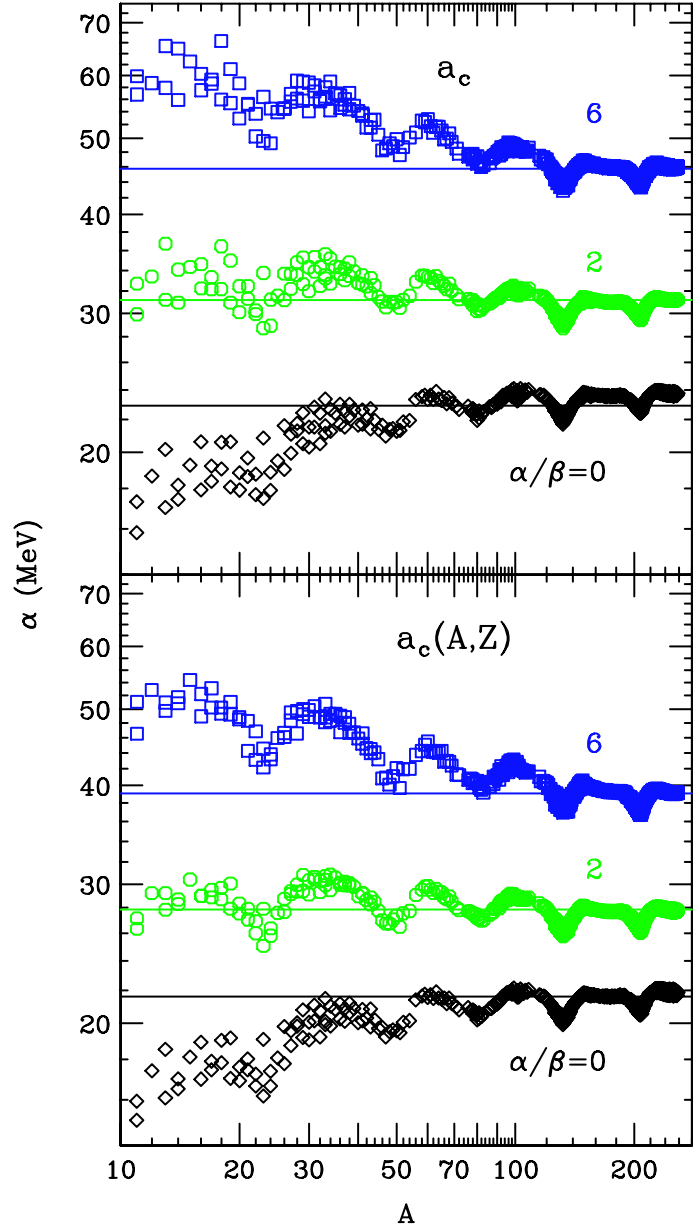}}
\vspace*{.2in}
\caption{
Values of the volume symmetry parameter $\alpha$ from inverting
the energy formula
at different values of $\alpha/\beta$,
for individual nuclei with $|N-Z|/A>0.2$,
as a function of $A$.  The top panel shows results for the
energy formula with $\Delta=0$ and the bottom for the formula
with  $\Delta$ given by Eq.\ (\protect\ref{delta}).  The
different symbols
represent results from the inversions for $\alpha/\beta = 6$
(squares), 2 (circles) and 0 (diamonds), respectively.  The
horizontal lines represent values of $\alpha$ from the
global fits at the different $\alpha/\beta$.
}
\label{fig:alphad}
\end{figure}

\begin{figure}
\centerline{\includegraphics[angle=90,
width=.40\linewidth]{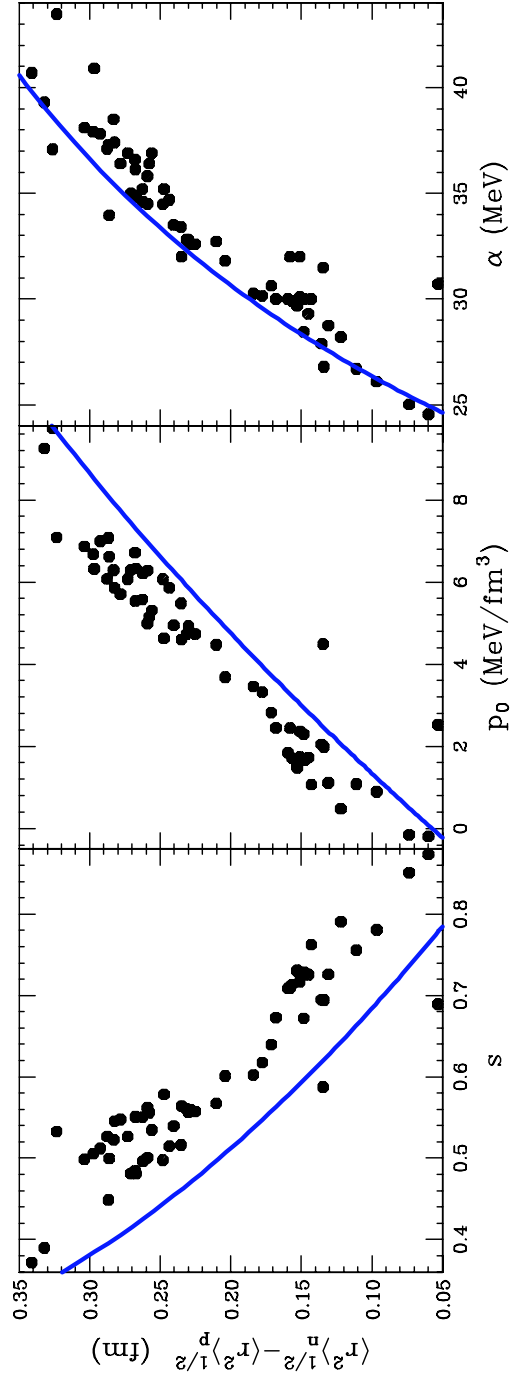}}
\vspace*{.2in}
\caption{
Size of the asymmetry skin in models vs the reduction factor $s$
for the symmetry energy at $\rho_0/2$ (left panel), vs the
scaled derivative
\protect\cite{oya98,fur02,bod03} of the energy
$p_0
= \rho_0^2 \, (dS/d\rho)$ at $\rho_0$ (center panel) and vs the
volume symmetry parameter $\alpha$ (right panel).
Symbols indicate results for the variety of mean-field models
explored by Furnstahl \protect\cite{fur02}.  Lines indicate
Thomas-Fermi results, from Eq.\ (\protect\ref{rmsnp}) using
(\protect\ref{abs}).
}
\label{fig:rrapn}
\end{figure}

\begin{figure}
\centerline{\includegraphics[angle=0,
width=.71\linewidth]{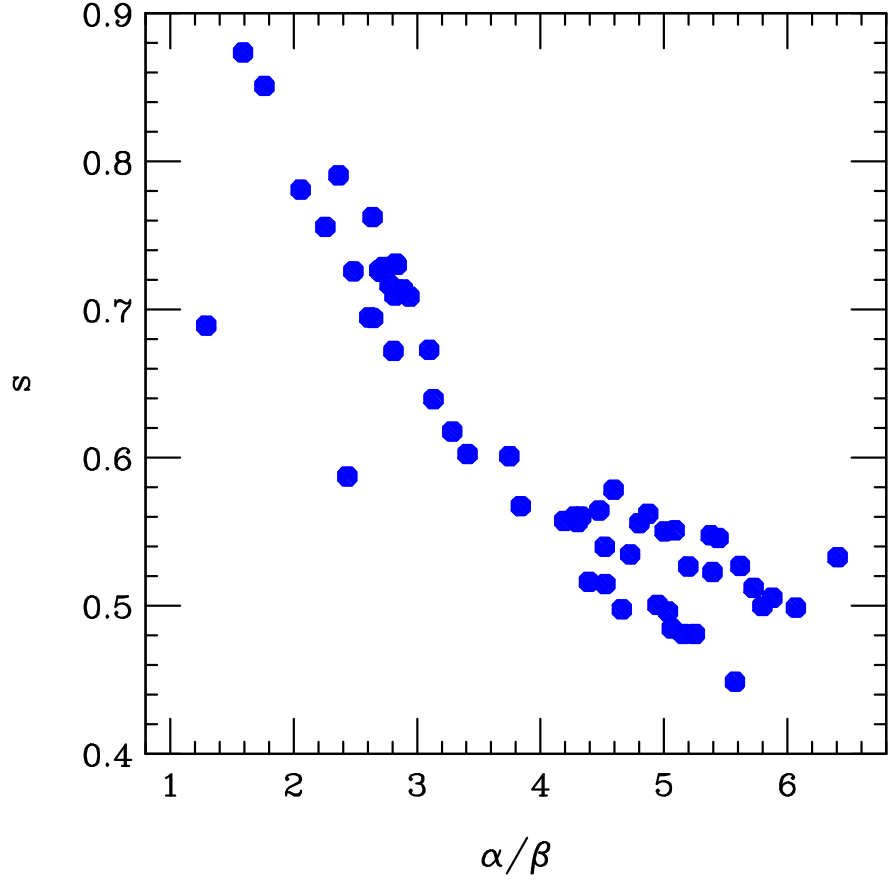}}
\vspace*{.2in}
\caption{
Reduction factor $s$
for the symmetry energy per nucleon
at
$\rho_0/2$
vs the symmetry parameter ratio $\alpha/\beta$ deduced
for the mean-field models
from the calculated \protect\cite{fur02} $^{208}$Pb skins and
Eq.~(\ref{rmsnp}).
}
\label{fig:sab}
\end{figure}

The hospitality extended by the Nuclear Theory Group of the
Tohuku University, where this work was initiated, is
gratefully acknowledged.  Acknowledged is further a
collaboration, on a topic related to the present work,
with Sergio
Souza and Betty Tsang \cite{sou03}.  Betty suggested an
early version of Fig.\ \ref{fig:alphad}.  A discussion with
Bill Lynch helped in clarifying an issue in Sec.\
\ref{sec:binding}.  Naftali Auerbach provided explanations
on giant resonances. Stefan Typel and Michael Thoennessen
helped in identifying literature.  Dick
Furnstahl generously supplied unpublished details of his
calculations \cite{fur02}, employed in Figs.\ \ref{fig:rrapn}
and \ref{fig:sab}.
A helpful discussion with Earle Lomon and a correspondence with
Aurel Bulgac are further also acknowledged.
This work was partially supported by the National Science
Foundation under Grants PHY-0070818, PHY-0245009 and INT-0124186.

\newpage

\newpage

\newpage

\newpage

\newpage

\newpage


%

\end{document}